\documentclass[12pt]{article}

\usepackage{latexsym,amsmath,amssymb,theorem,epsfig}

\usepackage{cite}
\usepackage{graphicx}
\usepackage{graphicx,subfigure}
\usepackage{epstopdf}
\usepackage[body={17.5cm, 21cm},right=2cm]{geometry}
\usepackage{amssymb}
\usepackage{amsmath}
\usepackage{bbold}
\usepackage{psfrag}
\usepackage{epsfig}
\usepackage{cancel}
 \allowdisplaybreaks[4]

\usepackage[all]{xy}

\DeclareFontFamily{OT1}{rsfs10}{}
\DeclareFontShape{OT1}{rsfs10}{m}{n}{ <-> rsfs10 }{}
\DeclareMathAlphabet{\mathscript}{OT1}{rsfs10}{m}{n}

\numberwithin{equation}{section}

\newcommand{\ns}{\normalsize}
\newcommand{\bra}{\langle}
\newcommand{\ket}{\rangle}
\newcommand{\cH}{{\cal H}}

\newcommand{\be}{\beta}

\newcommand{\x}{{\vec x}}
\newcommand{\xp}{{{\vec x}'}}
\newcommand{\y}{{\vec y}}
\newcommand{\p}{{\vec p}}

\newcommand{\kk}{{\vec k}}
\newcommand{\ad}{a^\dagger}
\newcommand{\kp}{{{\vec k}'}}

\newcommand{\mb}{\overline{M}}
\newcommand{\rr}{{\vec r}}

\newcommand{\ada}{a^\dagger}

\def\gsim{ \lower .75ex \hbox{$\sim$} \llap{\raise .27ex \hbox{$>$}} }
\def\lsim{ \lower .75ex \hbox{$\sim$} \llap{\raise .27ex \hbox{$<$}} }
\def\be{\begin{equation}}
\def\ee{\end{equation}}
\def\bea{\begin{eqnarray}}
\def\eea{\end{eqnarray}}

\begin{document}

\begin{titlepage}

\vspace{-5cm}

\title{
  \hfill{\ns }  \\[1em]
   {\LARGE New views on the \\ Low-Energy side of Gravity}
\\[1em] }
\author{
   Federico Piazza\footnote{fpiazza@perimeterinstitute.ca}
     \\
   {\ns Perimeter Institute for Theoretical Physics,
 Waterloo, N2L 2Y5, Canada}\\
\ns Canadian Institute for Theoretical Astrophysics (CITA), Toronto, M5S 3H8, Canada }

\date{}

\maketitle

\begin{abstract}
Common wisdom associates all the unraveled and theoretically challenging aspects of gravity with its UV-completion. However, there appear to be few difficulties afflicting the effective framework for gravity already at low energy, that  are likely to be detached from the high-energy structure. Those include the black hole information paradox, the cosmological constant problem and the rather involved and fine tuned model building required to explain our cosmological observations.  I review some on-going research that aims to generalize and extend the low-energy framework for gravity. In a quantum informational fashion, regions of space at a given time are treated and described as quantum subsystems, rather than submanifolds. The idea is to define a region of space through the quantum degrees of freedom of the matter fields ``living therein". I show how the correspondence sub-system/sub-manifold is realized in standard semi-classical gravity (``standard localization") and discuss the implications of alternative localization schemes. By exploiting further the subsystem description, I then consider the possibility that the usual GR metric manifold description might break down in the infra-red. This has implications on the description of the Universe on the largest scales and on dark energy.

\end{abstract}

\thispagestyle{empty}

\end{titlepage}

\nopagebreak
\small
\tableofcontents
\nopagebreak
\normalsize

\section{Introduction and Summary}

\subsection{A low-energy crisis?}
\label{sec:1}
According to the established paradigm, gravity is incorporated, together with -- and on the same foot as -- the other known interactions, in the low-energy framework described be the action
\begin{equation} \label{action}
S = \int \sqrt{-g}\left(M_{\rm Pl}^2 R + {\cal L}_{\rm matter}\right).
\end{equation}
In practice, at energies below $\sim$ 1 TeV,  ${\cal L}_{\rm matter}$ can be taken as the Standard Model Lagrangian plus some hidden and/or supersymmetric sector accounting e.g. for dark matter. It is widely believed that the theory (\ref{action}) can be trusted at three different levels:
\begin{itemize}
\item {\bf A} In the \emph{classical limit} (\ref{action}) reproduces classical General Relativity.
\item {\bf B} In the \emph{semiclassical gravity} regime the matter fields are quantized on a classical background spacetime and feed back into the Einstein equations through the expectation value of their renormalized stress energy tensor. 
\item {\bf C} Although non-renormalizable in the UV,  (\ref{action}) holds as a consistent low energy \emph{effective quantum field theory} \cite{cliff}. 
\end{itemize}
The classical limit has been tested with remarkable success in the dynamics of the solar system and in that of localized self-gravitating objects such as compact binaries etc\dots Points {\bf B} and {\bf C}, which are the natural quantum mechanical extensions of {\bf A}, still lack of experimental proofs. 

The above paradigm is known to face few difficulties:
\begin{itemize}
\item {\bf Problem 1} The \emph{black hole information loss paradox} appears already at the semiclassical level~\cite{haw} (and indeed in the low curvature/energy regimes nicely described in~\cite{peda} as ``solar system physics" limit): the quantized fields on a black hole background radiate to infinity through Hawking radiation. The black hole eventually evaporates and an initially pure state evolves into a thermal mixed state, which contradicts the apparent unitarity of \eqref{action}.
\item {\bf Problem 2} As an effective theory, (\ref{action}) produces, beside a series of marginal -- and practically unobservable -- gravitational operators, a \emph{cosmological constant} term~\cite{wein}, whose size is overestimated by several orders of magnitude. As far as a large cosmological constant can be considered a ``prediction" of \eqref{action}, such a prediction is wrong.

\item {\bf Problem 3} Only with some struggle does the above framework give account for our observations on the largest scales. It is true that there are models that can fit our cosmological data with impressive accuracy. However,  those explanations come with a cost: we need to assume two epochs of accelerated expansion (inflation and ``dark energy"), characterized by widely separated  mass scales, that require appropriate negative pressure components and, expecially for dark energy, a tremendous amount of fine tuning.
\end{itemize}
The common lore insists that the only problem with (\ref{action}) is its UV-completion. However,
the above difficulties appear already at the level of the low-energy description and, therefore, are likely to remain there regardless of the high-energy details of (\ref{action}). 
Not surprisingly, the candidate theories of quantum gravity look unable to address in a definite way any of the above.

Such a ``low-energy crisis" is not guaranteed to be anticipating a major reassessment of the current low-energy paradigm and might turn out to be overcome  through minor adjustments. The difficulty that especially seems to be calling for vigorous solutions is {\bf Problem 1}: in relation to it,
and to the more general framework of black hole thermodynamics~\cite{damour}, a violation of locality on macroscopic scales has been invoked~\cite{hol1,hol2,hol3,hol4,solu1,solu2,solu3,nonlocal1,nonlocal2,nonlocal3}. While there is general agreement that such a violation should correspond to highly entropic situations and/or to the presence of an horizon, it is not clear how the theory \eqref{action} should concretely be modified at low energy. Another strategy is to explicitly calculate transplanckian scattering amplitudes by including the effect of string-quantum gravity corrections (see~\cite{vene1,vene2,vene3} and references therein). While there is indication of production of soft quanta (gravitons) at the final stage of the collapse, the region of parameter space corresponding to a ``big" semiclassical black hole  seems very hard to analyze within this formalism. 

The remaining aspects of this crisis are milder  and cannot be considered as inconsistencies.  {\bf Problem 2} can be addressed, e.g., with  anthropic arguments \cite{wein1,bousso}. 
As far as {\bf Problem 3} is concerned, elaborate model building is not necessarily a bad strategy. If we look for inspiration, the physics of the last century provides two opposite examples. On the one hand, General Relativity explained Mercury perihelion with a deep theoretical breakthrough and an essential simplification. But the same did not apply, later on, to the complex phenomenology of particle physics: the Standard Model is a notable example of ``complicate model building", considerably fine tuned, and yet of extreme success and great predicting power. 

The possibility that the above difficulties might be addressed at once through a more radical breakthrough is however appealing and should also be explored in parallel with more standard approaches. As already argued, the problems at hand are likely to be detached  from what is usually meant by ``quantum gravity", i.e., the UV structure of General Relativity.  But what could be possibly changed in  the well constrained framework \eqref{action}?
Local quantum field theory does not allow for much freedom: save for a cosmological constant, the Einstein-Hilbert term $\sqrt{g} R$ 
is the leading operator compatible with general covariance.   We are left with the alternative of treating gravity on a different foot than the other interactions\footnote{Although this is not the dominant point of view, the ``anomaly" of some aspects of gravity has been occasionally pointed out.  Dyson~\cite{dyson} (see also the follow-up works~\cite{br}) has questioned whether  the concept of graviton has any operational definition i.e. whether any conceivable physical process is able to detect a single graviton. Other anomalous aspects of gravitational interactions are surveyed by Zee in the thought-provoking contribution~\cite{zee}. The entire emergent gravity program (see, among others,~\cite{em1,em2,em3,em4} and references therein) effectively invests in the hypothesis of treating gravity differently from the very beginning. The idea of gravity as a macroscopic entropic force and of spacetime as emergent has also been strongly advocated in~\cite{verlinde}. Also the views expressed in \cite{olaf,fotini} are very close to the original motivations of the present analysis.}.

The purpose of this note is to review a direction of ongoing research~\cite{fedo1,fedo2,fabio1,fabio2,sergio,ultrastrong,ircompletion} that aims to generalize  the semiclassical regime of \eqref{action} (point {\bf B} above). The main idea is, instead of dealing with spacetime and regions of space, trying to deal directly with the quantum degrees of freedom of the matter fields and therefore with the more general concept of \emph{quantum subsystem}. 
The basic observation, or prejudice, at the basis of this research is that spacetime relations are operationally defined with the matter fields (rods, clocks, particle detectors etc\dots) and that therefore any region of space is described, before anything else, by the quantum degrees of freedom of $\cal L_{\rm matter}$ ``living therein"~\cite{fedo1,fedo2}. 

The picture in standard semiclassical gravity is reviewed in Sec. \ref{2}. 
In that regime regions of space at a given time (``this room, now") have a dual description: they are \emph{submanifolds} for GR but they are also \emph{subsystems} as they ``contain" the local quantum degrees of freedom of the matter fields. As I am going to review in detail, 
the correspondence submanifold/subsystem is set by the local operators of the field theory~\cite{fabio1,sergio}. 
This duality submanifold/subsystem is not much emphasized usually, but is often used implicitly. For instance, in standard calculations of vacuum entanglement entropy~\cite{sergio,en1,en2,en3,en4,en5}, regions of space are effectively treated as quantum subsystems and the degrees of freedom 
external (the ``environment" in the language of quantum information) to the region considered are traced over in order to compute reduced density matrices.
In Sec. \ref{2} it is also shown with a simple example in which sense the \emph{subsystem-description} is more general than the submanifold-one. In the rest of the paper I exploit the \emph{subsystem-description} as a mean to extend/modify the present framework \eqref{action}. In Secs. \ref{sec3} and \ref{sec4} I will indicate two different -- but not incompatible -- strategies that have been the subject of a recent series of papers.

\subsection{Ultra-Strong Equivalence Principle: a large distance modification}

The first strategy~\cite{ultrastrong,ircompletion}, developed in Sec.~\ref{sec3}, fully exploits the \emph{subsystems-description}, by considering the possibility that the \emph{submanifold-description} might break down on large scales. In doing this, I try to address directly {\bf Problems 1} and {\bf 2} above. In which sense are those problems  related? Both difficulties can be associated with  the presence of spacetime dependent terms in the vacuum expectation value (VEV) of the (bare) energy momentum tensor. Schematically, for a massless field in a spatially homogeneous spacetime, the energy density VEV can be expanded at  high momenta as 
\begin{eqnarray} \label{structure}
\langle T_0^0(x,t)\rangle_{\rm bare} &=& \int d^3 k\left(k  + \frac{f_{\rm quad}(t)}{k} + \frac{f_{\rm log}(t)}{k^3} + \dots  \right) \\[2mm]
& = &\ \  {\rm local\ terms}\ \ +\ \ {\rm non \ local\ terms}. \nonumber
\end{eqnarray}
There is a standard procedure~\cite{fulling,fulling2,birrell} to make sense of the infinities appearing in \eqref{structure}. The local terms can be subtracted by local gravitational counterterms (cosmological constant, Newton constant and higher order local operators). The finite non-local terms represent the genuine particle/energy content of the choosen ``vacuum" state. 

Now imagine a different theory, where the $f$s are just absent from the beginning. Such a theory is going to be similar to standard semiclassical gravity at small distances (small compared to the local curvature), as the time-dependent pieces are effectively infra-red (IR) with respect to the quartically divergent term. The terms responsible for black hole evaporation, and its puzzling consequences ({\bf Problem 1}), would just be absent. 
Moreover, bypassing the usual procedure of stress tensor renormalization might, at the same time, shed some new light on {\bf Problem 2}, at least at the level of non-quantized semi-classical gravity. If, in the IR-completed theory,
all the time dependent terms just do not exist, then we do not need to renormalize the stress tensor anymore. Of course, we are still left with the leading (constant) quartic divergence, but we can live with it and treat it, as we do in flat space, by normal ordering.   Finally, it is interesting to reabsorb the  quadratic divergence in the IR, rather than with a local counterterm, because the required modification is of the right order of magnitude to give interesting cosmological implications. 

The above reasoning suggests to consider a ``Ultra-Strong" version of the Equivalence Principle, namely, \emph{For each matter field or sector sufficiently decoupled from all other matter fields, there exists a state, the ``vacuum", for which the expectation value of the (bare) energy momentum tensor reads the same as in flat space, regardless of the configuration of the gravitational field.} In Sec. \ref{sec3} I explore the conjectured theory by ``IR-completing" standard semi-classical gravity, i.e., with a modification of the field operators in the infra-red. 
Analogously to other modifications of field operators that have been proposed in the literature (typically in the UV), the present one seems to suggest a breakdown (here, in the IR, at large distances) of the metric manifold description. 

\subsection{Localization: a short (but not so short-) distance modification}

The second strategy, described in Sec.~\ref{sec4}, does not imply a breakdown of the ``manifold description" and has to deal with \emph{localization} \cite{fabio1,fabio2,sergio} i.e. which quantum degrees of freedom of the matter fields should be associated with a given region of space. This boils down to ask how the correspondence submanifold/subsystem should actually be realized. 
The low-energy framework \eqref{action} implicitly contains the standard answer to this question: to a given region of space at time $t$ correspond the local relativistic fields $\phi({\vec x},t)$ with labels ${\vec x}$ therein defined. This is the standard localization scheme. In Sec. \ref{sec3} I give motivations for -- and explore some of the consequences of -- using a different localization scheme, related to the Newton Wigner (NW) position operator~\cite{nw,stefa}. When written in the Newton Wigner basis, the Hamiltonian of a matter field theory is not of the local form $H = \int d^3 x {\cal H}$ but is expressed as a multiple integral. For a free scalar field in Minkowski spacetime
\begin{equation} \label{nwhamiltonian1}
H \ = 
\int  d^3x \, d^3y \, K(|{\vec x} - {\vec y}|) \, a^\dagger({\vec x}) a({\vec y}), 
\end{equation}
where $a({\vec x})$ are operator localized according to NW and $K$ a kernel that 
dies off as $K \sim e^{- m|{\vec x} - {\vec y}|}$ 
for $|{\vec x} - {\vec y}|\gg m^{-1}$ where $m$ is the mass of the field and $K \propto |{\vec x} - {\vec y}|^{-4}$ in the massless case. Within the matter sector, the dynamics between asymptotic states is trivially unaffected as I am just rewriting the same operator $H_{\rm matter}$ on a different basis: different localizations give
the same cross sections, decay rates etc\dots 
What NW-localization implies is a different fine-grained space-time description. Therefore, when gravity is also included in the game, deviations from \eqref{action} are suggested. 
Imagine we want to calculate the average energy -- and the contribution to the gravitational field -- of a certain region of space. 
According to NW the corresponding operator is obtained from \eqref{nwhamiltonian1} by limiting both the integration variables to the region considered. 
For regions of space much bigger than $m^{-1}$ and/or the inverse temperature of the state considered, the average energy is the same as in standard localization. But for smaller regions the average energy is non extensive and somewhat smeared on the microscopic scales. This suggests that, as far as the coupling gravity-matter is considered, \eqref{action} should be trusted only in some macroscopic limit i.e. at distances larger than Compton wavelengths/inverse temperatures. While this should not contradict all known GR tests, there is a relevant amount of non-locality inherited by gravity at microscopic, and yet unexpectedly large, distances.


\section{Regions of Space as Quantum Subsystems} \label{sec:2}

\subsection{Breaking a quantum system into pieces} \label{sec:2.1}

It is interesting -- and rather counterintuitive -- the description that quantum mechanics provides for the partition of a system, the division of a quantum system into ``pieces". Most often, situations are considered in which the division into parts is already given. Typically, we are provided with two systems, $A$ and $B$ (for example, two different particles in first quantization), described by the two Hilbert spaces $\cH_A$ and $\cH_B$ respectively, and we have to describe the composite system of the two. The quantum degrees of freedom of such a system are known to live in the tensor product Hilbert space $\cH = \cH_A \otimes \cH_B$. Tensor products are most often used to construct the Hilbert space of a composite system once the spaces of the components are known. However, one may also go the other direction
and see in how many ways a given Hilbert space can be decomposed into ``virtual subsystems" \cite{paolo1} or, in other words, in how many ways a Hilbert space can be given a \emph{tensor product structure} (TPS).

The partitions of a quantum system -- i.e. all possible ways that a quantum system can 
be divided into ``parts'' -- have a mathematical structure significantly different from, for example,
the partitions of a set in set theory. In set theory you can choose a bi-partition
$A-B$ by going through each element of a (countable) set and deciding whether it belongs to subset $A$ or $B$.
Clearly, finite sets admit only a finite number of possible partitions. Analogously, 
a finite lattice can be divided into sub-volumes in a finite number of ways. 

Let us consider the extreme case of a lattice made of just two-spins, two qubits: $A$ and $B$.  As a set of points/events (manifold-description), this very trivial lattice can be partitioned only in one way: the site $A$ on one side and the site $B$ on the other. 
However, a two-qubits quantum system has Hilbert space $\cH = \mathbb{C}^4$. So, asking in how many ways the quantum system (\emph{subsystem-description}) can be divided boils down to ask in how many ways we can decompose $\mathbb{C}^4$ into a tensor product $ \mathbb{C}^2_A \otimes \mathbb{C}^2_B$ (indexes $A$ and $B$ identify each of the two identical components $\mathbb{C}^2$).
The answer is: infinite. In fact, given any orthonormal basis
$\{|a\ket,\, |b\ket,\, |c\ket,\, |d\ket\}$ of $\mathbb{C}^4$,  one way of partitioning the quantum system is through the identification 
\begin{equation}\label{identification1}
|a\ket \simeq |0\ket_A \otimes |0\ket_B, \quad |b\ket \simeq |0\ket_A \otimes |1\ket_B, \quad |c\ket \simeq |1\ket_A \otimes |0\ket_B, \quad |d\ket \simeq |1\ket_A \otimes |1\ket_B, 
\end{equation}
where $\{|0\ket_{A}, \, |1\ket_{A}\}$ and $\{|0\ket_{B}, \, |1\ket_{B}\}$ are some choosen basis in $\mathbb{C}^2_{A}$ and $\mathbb{C}^2_{B}$ respectively. 
A different partition is defined by the choice of another orthonormal basis, say $\{|a'\ket,\, |b'\ket,\, |c'\ket,\, |d'\ket\}$, to use for the one to one correspondence (\ref{identification1}). All possible partitions of $\mathbb{C}^4$ are thus given by the elements of the group $SU(4)$ except that, within $SU(4)$, there 
are also transformations that merely correspond to a change of basis in either of the two factors $\mathbb{C}^2$. These transformations have to be factored out since they don't change the partition, leaving us with the group $SU(4)/SU(2)^2$: those are all the inequivalent ways we can separate a two-spin\footnote{More generally, a $d^N$-dimensional Hilbert space  can be partitioned into $N$ smaller systems each of dimension $d$ and such partitions are in one to one correspondence with the elements of  $SU(d^N)/SU(d)^N$} system! Note therefore that
quantum degrees of freedom, even when finite, can be split in an infinite number of ways.
Not only can you choose whether some of them belong to, say, subsystem $A$ or $B$, but, as opposed to the elements of a set or the sites of a lattice, you can unitarily mix them before the splitting, in such a way that they completely lose their individual identities~\cite{paolo1}. 

More practically, a partition of a quantum system is assigned by specifying a set of 
accessible observables \cite{paolo2}. Consider, in fact, a system that is already divided into two parts. This time I will call them  $P$ and $R$. In fact, since I will apply these arguments to spacetime, the letter $P$ stands for \emph{Place} and $R$ for \emph{Rest} (of the system):  $\cH = \cH_P \otimes \cH_R$. Say that I have two sets of observables, ${\cal A}^j_P$ and ${\cal A}^k_R$,  
separately defined in subsystem $P$ and $R$ respectively. The indexes $j$ and $k$ just label the different observables in the two sets. It is trivial to define new observables, this time on the entire system,  by extending those two sets as follows,
\begin{equation}
{\cal A}^j_P\  \longrightarrow\ {\cal A}^j(P) \ \equiv \ {\cal A}^j_P \otimes \mathbb{1}_R, \quad \qquad
{\cal A}^k_R\  \longrightarrow\ {\cal A}^k(R) \ \equiv \ \mathbb{1}_P \otimes {\cal A}^k_R .
\end{equation}
In practice,  from the observable ${\cal A}^1_P$ acting on $P$, I am building a new observable ${\cal A}^1(P)$, defined on the entire system, that acts as ${\cal A}^1_P$ on $P$ and trivially  (as the identity) on $R$. By construction, 
\begin{equation}\label{commute}
[ {\cal A}^j(P), {\cal A}^k(R)]\, =\, 0.
\end{equation}
The idea here (see \cite{paolo2} for more details) 
is to use this simple argument the other way around and to define a partition starting from the observables.
That is, if I manage to isolate two subalgebras ${\cal A}(P)$ and ${\cal A}(R)$, within the algebra of observables acting on $\cH$, satisfying \eqref{commute}, then they induce a unique\footnote{More precisely, that is true only if 
the two subalgebras generate the entire algebra of operators on $\cH$ \cite{paolo2}}  
bipartition $\cH = \cH_P \otimes \cH_R$.
Since in quantum field theory (QFT) the usual local observables commute at space-like separated events, 
we have a straightforward realization of \eqref{commute} and we can use local fields to define 
a local TPS at each time $t$.

\subsection{Regions of space in standard semiclassical gravity} \label{sec:2.2}

\begin{figure}[t] 
\vspace{-1.5cm}
\includegraphics[width=38pc]{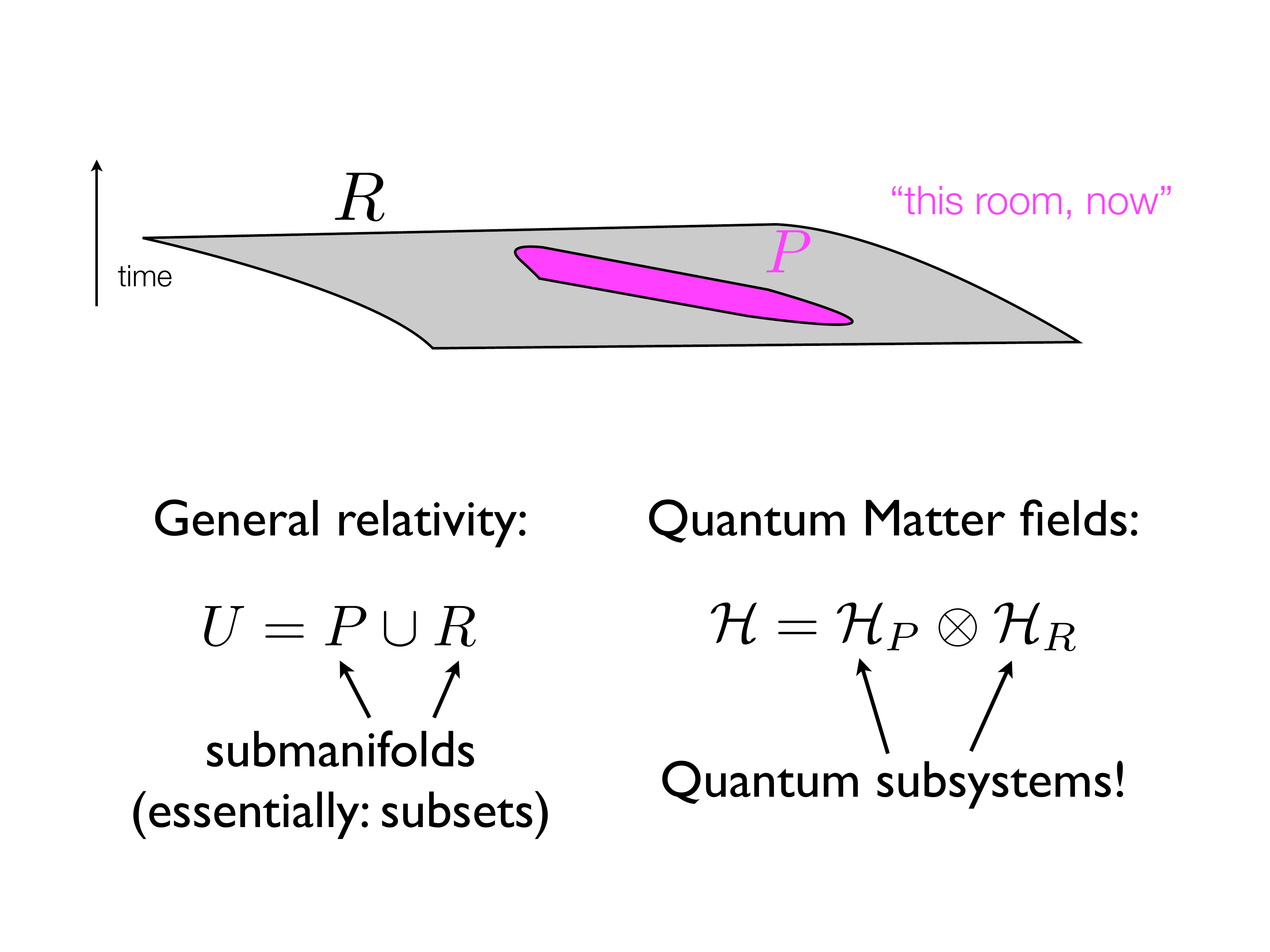}
\vspace{-.5cm}
\caption{\label{figure} Schematic representation of the two complementary descriptions of a region of space at a given time in semiclassical gravity. For general relativity ``this room now" is a submanifold with boundary. For the quantized fields ``this room now" is a quantum subsystem}
\end{figure}

In the standard formulation of semi-classical gravity, matter fields are quantized on a curved background manifold and feed back into Einstein equation through the average of  their renormalized energy momentum tensor. Consider a spacetime with a global time foliation labeled by a time parameter $t$. The Universe $U$ as a whole at time $t$ is a three-dimensional manifold in the GR description. The matter quantum fields are instead described by a quantum state living in a Hilbert space $\cH$. Now consider a region of space $P$ at time $t$ (``this room, now"), that has two complementary descriptions \cite{fedo1,fabio1,sergio} schematically depicted in Fig. \ref{figure}: it is a \emph{submanifold} according to GR and a \emph{quantum subsystem} for the quantized fields. 
Here the letter $P$ stands for \emph{Place} and $R$ will refer to the \emph{Rest} of the Universe. Therefore, the entire system $U$, in GR, is given by the union $U = P \cup R$.
From the point of view of the quantized fields, the same division is a quantum partition $\cH = \cH_P\otimes \cH_R$, a tensor product
decomposition of the total Hilbert space $\cH$ of the field theory. 
In calculations of entanglement entropy~\cite{sergio,en1,en2,en3,en4,en5} it is common to consider regions of space as quantum subsystems and trace e.g., over the quantum degrees of freedom ${\cal H}_R$ of the ``environment" $R$.

The correspondence sub-manifold/sub-system is explicitly realized by  the set of local operators $A(x^i,t)$ of the field theory. Those act on the quantum system $\cH$ but have labels $x^i$ living on the three-dimensional manifold $U$. As a consequence, we can take integrals of scalar local operators over some region, which are still operators acting on the total Hilbert space $\cH$. Schematically, 
\begin{equation} \label{bah}
{\cal A}(P,t) = \int_P d^3 x \sqrt{-g^{(3)}}\ A(x^i, t) ,
\end{equation}
where $g^{(3)}_{i j}$ is the three dimensional metric on the fixed time hypersurface and integration has been limited to submanifold $P$. By integration over $R$, an element of the algebra ${\cal A}(R,t)$ can be built in a similar way. 
Because of the commutation relations between the local fields, \eqref{commute} is satisfied.
Again, this is just saying that ${\cal A}(P,t)$  acts non-trivially only on the quantum subsystem $\cH_P$ corresponding to the region of space $P$ that has been integrated over, and as the identity on the rest of the system $\cH_{R}$. On the basis of the discussion above, it is now clear that the algebra of operators ${\cal A}(P,t)$ determines \cite{paolo2} the partition $\cH = \cH_P \otimes \cH_{R}$ of the quantum system/Universe and therefore can be taken as a definition of the region of space $P$.

\subsection{Aside: spacetime from scratch (an attempt)}

Defining regions of space as quantum subsystems seems a useless complication in the standard formalism \eqref{action}, because the correspondence  subsystem/submanifold is always automatically at work: such a correspondence is set once and for all by \eqref{bah}, once we are given a complete set of local operators $A(x^i,t)$. The main focus of this paper is that of using the \emph{subsystem description} as a tool for generalizing and extending  \eqref{action}. The two main strategies proposed are reviewed in Sec. \ref{sec3} and Sec. \ref{sec4}.
However, considering regions of space as quantum subsystems also offers the opportunity for a more speculative/philosophical attempt that I will mention in this subsection. Namely, trying to recover spacetime relations -- and spacetime itself -- on theoretic-informational grounds, from the mutual relations between matter fields degrees of freedom. I will refer the reader to \cite{fedo1} for more details. See also \cite{olaf,fotini} for analogous points of view.

The basic observation is that 
any spacetime measurement is made by the mutual relations between objects, fields, particles etc\dots\  Therefore, any operationally meaningful assertion about spacetime is  intrinsic to the degrees of freedom of the matter (i.e. non-gravitational) fields; concepts such as ``locality'' and ``proximity'' should, at least in principle, be operationally definable entirely within the dynamics of the matter fields. In this respect, the usual approach of QFT follows quite an opposite route: the fundamental degrees of freedom are there labelled, from the beginning, by the spacetime points and locality is given \emph{a priori}. 

The goal of such a ``spacetime reconstruction program" is to recognize, among all the possible quantum partitions (TPS) of the matter fields Hilbert space,  
\begin{equation} \label{deco}
\cH_{\rm Universe} \ = \ \cH_A \otimes \cH_B \otimes \cH_C  \otimes \dots \ ,
\end{equation}
those partitions that characterize the \emph{localized} degrees of freedom. In other words, to recognize the partitions whose factors ($\cH_A$, $\cH_B$ etc\dots) correspond to regions of space.
A decomposition into subsystems such as (\ref{deco}) can in fact be assigned without any reference to the locality or to the geometric properties of the components. Consider, for example, the decomposition in subsystems of given momentum: that is usually a very poorly ``localized" decomposition. 

Consider first an arbitrarily assigned TPS on a Hilbert space. That does not have much significance on its own: without any observable/operator of some definite meaning it 
looks impossible to extract any physical information about the system. However,
by knowing the dynamics inside $\cH_{\rm Universe}$, i.e. the unitary
operator $U(t_2,t_1)$ that, in the Schroedinger representation, evolves the state 
vectors according to $|\Psi(t_2)\ket = U(t_2,t_1)|\Psi(t_1)\ket$, we can, at least, follow 
the evolution of the \emph{correlations} between the subsystems. Correlations play a central role in Everett's view of quantum mechanics. In his seminal 
dissertation \cite{everett}, the relation between quantum correlations and mutual information is deeply exploited and measurements are consistently described as appropriate unitary evolutions that increase the degree of correlation between two subsystems: the ``measured'' and the ``measuring''. By taking Everett's view (see also Rovelli's ``relational" view~\cite{rovelli}), one can try to re-interpret the evolution of the system $|\Psi(t)\ket$ 
as measurements actually going on between the different parties $A$, $B$, $C$ etc\dots. It is  
compelling that locality itself and the usual local observables of direct physical interpretation may be eventually picked out within such an abstract scheme.

So the problem is trying to characterize, from the development of the correlations between the factors in \eqref{deco}, the class of TPSs that single out localized subsystems, the ones associated with regions of space.
As a source of correlation/information I will consider quantum entanglement, which, 
for a bipartite system $AB$ in a pure state $|\Psi\ket$, is measured
by the \emph{von Neumann entropy} $S(A) = - {\rm Tr}_A (\rho_A \log_2 \rho_A)$, 
where $\rho_A = {\rm Tr}_B |\Psi\ket\bra\Psi|$.
The most elementary type of spacetime relation that one can try to define intrinsically from
the dynamics of general subsystems is that of \emph{mutual spacetime coincidence}, what may be intuitively
viewed as ``being in the same place at the same time" or just ``having been in touch''. 
Inspired by the known local character of physical laws, we attempt to define 
coincidence by means of physical interactions, i.e. to define  two parties as 
``having been coincident" if they ``have physically interacted" with each other. 
By choosing the production of entanglement as a ``measure'' of interaction 
a sufficient condition for spacetime coincidence can be given:
\begin{quote}
\emph{Spacetime coincidence (sufficient condition)}: If before the instant $t_1$ the subsystems $A$ 
and $B$ are in a pure state \emph{i.e.} $S(A;t<t_1) = S(B;t<t_1) =0$ and, at a later time 
$t_2$ they are entangled, $S(A;t_2) = S(B;t_2) > 0$, without, during all the process, having mixed with anything else,
$S(AB;t<t_2) = 0$, then $A$ and $B$ \emph{have been coincident} with each other between $t_1$ and $t_2$.
\end{quote}
As a sufficient condition, the one above stated is rather strict: 
there are a number of physical situations that one would legitimately consider as ``spacetime coincidence relations'' but do not fit into the above definition. Most notably, two systems may interact with each other
while being already entangled with something else; i.e. not initially in a pure state. 
In this case, however, it is hard to give a quantitative definition of contiguity because a reliable measure of entanglement for multipartite systems is still a matter of debate. 

Note that, if $A$ and $B$ have been coincident, coincidence generally applies also to many 
other ``larger'' systems containing $A$ and $B$ as subsystems. 
The opposite can also be true. Say that $A$ is itself a composite system \emph{i.e.} $\cH_A = \cH_{A1}\otimes\cH_{A2} \otimes \dots$; we may discover that the sub-subsystem $A2$ is in fact ``responsible'' at a deeper level for the coincidence between $A$ and $B$. 
Again, the smaller the dimension of the systems which coincidence is recognized to apply, 
the more finely grained the spacetime description we are able to give. 

The notion of coincidence can be applied to subsystems belonging to 
arbitrary TPSs and therefore possibly maximally unlocalized, such as, in ordinary QFT in flat space, those associated with the modes of given momentum. In the TPS of the localized subsystems, however, during an infinitesimal lapse of time, each subsystem creates new correlations with the smallest possible number of other subsystems: its ``neighbors''.  We argue therefore the following generic property of the TPSs that single
out localized subsystems: 
\begin{quote}
\emph{Locality Conjecture:} 
``Localized subsystems'' have the minimum tendency to create \emph{coincidence} relations with each other: the tensor product structure that singles out \emph{localized systems} is the one in which the entanglement of initially completely factorized states \emph{minimally} grows during time evolution. 
\end{quote}
In \cite{fedo1} generic interacting second quantized models with a finite number of 
fermionic degrees of freedom have been considered. The symmetries of the Hamiltonian (in this case the conservation of number of particles) dramatically restrict the possible TPS choices. By applying the above conjecture to a one-dimensional 
Heisenberg spin chain and to two particles states the tensor product structure usually associated 
with ``position'' is recovered.

\section{The Ultra-Strong Equivalence Principle} \label{sec3}

The first strategy~\cite{ultrastrong,ircompletion} fully exploits the \emph{subsystems-description} of the regions of space by considering the possibility that the \emph{submanifold-description} might break down in some regimes. A breakdown of spacetime continuum is sometimes advocated in relation to the physics at the Planck scales. However, Planck scale modifications might not be able to address the problems that afflict gravity at low energy, such as those related in Sec. \ref{sec:1}. If, as expected, the decoupling of the highest modes is at work, then there are poor chances that those difficulties might be resolved by working on the UV structure of \eqref{action}.  This is the main motivation for considering IR-modifications instead.

\subsection{IR-modified gravity with no new mass scale.}

The IR-modifications of gravity considered so far (see, e.g. \cite{dgp,dgp2,dimo,justin,justin2} and the nicely written review~\cite{dvalirev}) include the presence of a new mass scale in the theory
and are often effectively equivalent, in some appropriate sense, to giving the graviton a tiny mass. These models have been studied in relation to dark energy and to the cosmological constant problem. Interestingly, by changing the dynamics of the gravitational field at large distances, a mechanism of self-acceleration of the universe and/or of filtering for the
cosmological constant's zero mode can be provided. Those approaches modify the dynamics of gravity and leave intact its basic geometrical description. 

If we are to look at \eqref{action} as a small-distance approximation, it seems in fact very natural to follow the massive gravity approach and introduce a new scale in the theory, corresponding to the threshold below which departures from GR  become relevant~\cite{orfeu}. However, by taking GR itself as an example, one can also argue for a logical  alternative to that. In fact, GR does not contain any more mass parameter than Newtonian gravity. Moreover, GR can also be considered as an extremely successful attempt of IR modification, as it describes gravity as something 
effective only at sufficiently (parametrically) large distances, i.e. \emph{outside}  Einsten's free-falling elevator. According to GR, what breaks down in the IR
is the flat-space/non-gravitational physics approximation, which is only valid for systems much smaller than the inverse curvature. In other words, the IR scale of GR is not a new parameter, but is dynamically set by the curvature. 
This can be seen, for instance, in three dimensions, by considering the area of a two-sphere of radius $l$ and volume $V$ in presence of curvature. Such an area receives corrections from the flat-space expression of the type
\begin{equation} \label{1}
A(l) \ = \ 4 \pi l^2 \, (1 + {\cal O}(l^2 R)) \ =\  (36 \pi V^2)^{1/3}\, (1 + {\cal O}(R V^{2/3}))  .
\end{equation}
In order to modify GR without new mass scales, what seems to be required is a further curvature-dependent
subleading effect that systematically modifies the geometrical description of GR at large distances. I am thinking about a generalization of the Riemannian metric manifold: a geometrical object that resembles the curved space of GR in the vicinity of each point/event in the same way as the curved space of GR -- locally -- resembles flat space.

Unfortunately, I am not able, at present, to provide a general and mathematically consistent framework for such a modification. 
It is known that there is a strict connection between the geometric properties of a manifold and the spectrum of the differential operators \cite{barv,heat} or the algebra of functions \cite{connes} therein defined; such abstract characterizations have already been used for generalizing common geometrical concepts and the description of spacetime itself~\cite{connes,sw}. However, so far, attempts in this direction have always been applied to the UV and intended to modify spacetime at the smallest scales. Although perhaps less intuitive, a modification on large scales may be able to address more directly some of the problems described in the introduction that trouble the low energy framework of gravity. 

Spectral geometry provides some useful insights for generalizing common  geometric concepts. A differential operator on a Riemannian manifold (e.g. the Laplacian on a sphere) has a set of eigenvalues that can be studied, e.g., by heat kernel expansion techniques~\cite{barv,heat}. The idea now is to change the asymptotic behavior of such eigenvalues in the IR limit. Arbitrary spectral modifications may not always be supported by a corresponding metric manifold. Other frameworks for generalizing Riemannian manifold are Finsler geometry~\cite{finsler} (see~\cite{finslerDE} for applications to Dark Energy) and torsion gravity~\cite{torsiongravity,torsionlimit}.
As a simple guiding principle  
for such a modification, I propose is a stronger (or quantum) version of the equivalence principle. While its main motivations are the set of problems described in the introduction, some more general and philosophical motivations are given in the following section.

\begin{center}
\begin{figure}[t] 
\vspace{-1.8cm} 
\hspace{.2cm}
\includegraphics[width=38pc]{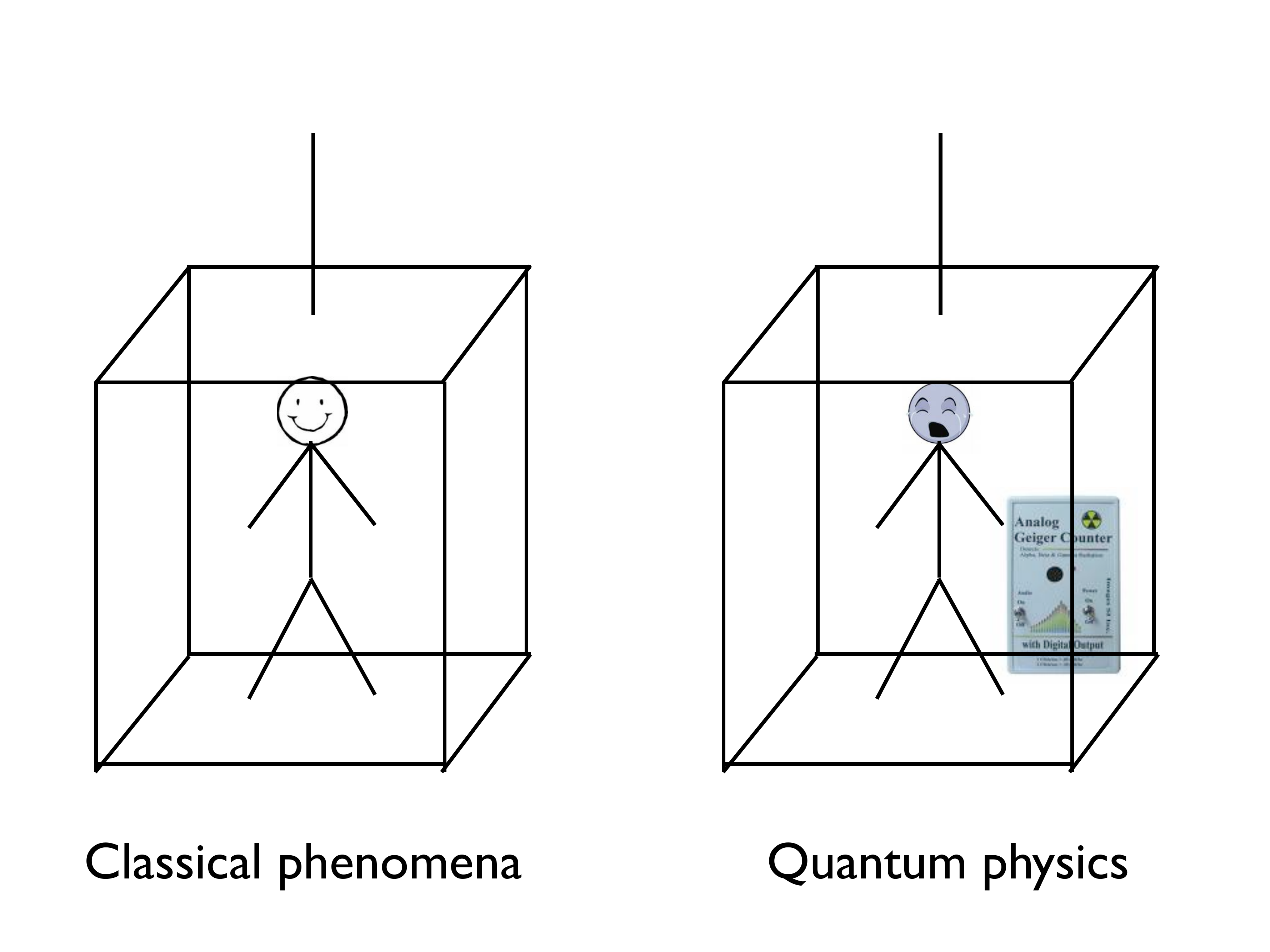}
\caption{\label{figure2} Non-local gravitational effects contribute to the (local) energy momentum tensor.}
\end{figure}
\end{center}

\subsection{The Equivalence Principle: Gravity as an IR effect.}\label{sec:3.2}

The equivalence principle (EP) can be formulated very simply: \emph{inside a sufficiently small free-falling elevator you do not see the (classical) effects of gravity}.  Among the many celebrated implications of general relativistic physics, the view that I aim to stress in this Section is that EP forces us to consider and describe gravity as an IR phenomenon, whose effects are visible only outside the free-falling elevator. How EP turned into a consistent theory is well known: gravity is beautifully encoded in GR as the geometry of the physical space-time and therefore its effects are automatically suppressed within those systems that are much smaller than the inverse curvature. 
By changing (curving) the large-scale structure of spacetime, GR leaves the smallest systems free of classical gravitational effects. 

Gravity was originally banned from the free-falling elevator but its effects reappeared, after the developments of quantum theory, through the ``back door".
The fields quantized on a curved manifold are sensitive to the global properties of spacetime because their modes are defined on the whole of it. A well known and understood consequence of this is particle creation~\cite{birrell}. 
Can we detect such an effect inside the free-falling elevator with e.g. particle detectors? 
It is arguable that a small system cannot detect quanta of arbitrarily small energies. However, while that is surely generically true in practice, there seem to be no universal and in principle argument that limits the possibility of particle detection by small detectors. Let us think of a particle detector as a physical system of size $L$ that ``detects" a quantum through an internal transition to an excited state.
Within a system of size $L$, the first excited level is typically at energy $1/L$ above the ground state and, therefore, cannot be excited by photons of much lower energies. However, the structure of the higher energy meta-stable states is often very rich, and energy differences among such states can be arbitrarily small. An atom, for example, has extremely fine energy transitions that, in principle, allow to ``detect" photons of wavelengths much larger than its size. In summary, it looks difficult to think of a definite no-go theorem for detection based only on the size of the detector and the energy of the quantum and independent of the details of the dynamics internal to matter field theory. On the opposite, there is a very precise and quantitative sense in which tidal forces in GR become small within small systems. 

But anyway, the question whether particle production challenges the spirit of EP is debatable and mostly a matter of terminology.  Still, it is tempting to try a different path than the one historically followed: rather than imposing field quantization on top of a curved manifold, try to upgrade the equivalence principle, make it ``ultra-strong" and extend it to the quantum phenomena! 
Thus, I will consider a stronger version of EP, in which all the effects of gravity are definitely forbidden inside the elevator, including the quantum effects that in the standard semi-classical treatment lead to particle creation. Operationally, such a principle can be stated as
\emph{for each matter field or sector sufficiently decoupled from all other matter fields, there exists a state, the ``vacuum", that is experienced as empty of particles by  each free-falling (inertial) observer.}

The above conjecture is adventurous. Hawking radiation and quantum particle production are extremely well understood aspects of the standard low-energy framework~\eqref{action} and, by now, are part of theoretical physicists' intuition and understanding. Any experimental evidence for these phenomena would be an indirect confirmation of the low-energy framework~\eqref{action}. 
Inflationary models beautifully exploits the mechanism of quantum particle creation  in order to produce the primordial spectrum of  cosmological fluctuations. However,
the present knowledge of the earliest Universe is too model dependent (and largely based on the assumption that such a mechanism is at work!) for that to be considered a proof. Before our hints turn into clear experimental evidences it is healthy to keep an eye out for alternatives.

In order to recover the information from an evaporating black hole, modifications  of ~\eqref{action} have in fact been proposed, in essentially non-local,  holographic, directions~\cite{solu1,solu2,solu3, nonlocal1, nonlocal2,nonlocal3}. It is argued that the local description~\eqref{action} is brutally wrong 
in highly entropic situations and/or in the presence of horizons/closed trapped surfaces. In parallel with those insights, it looks natural to consider the (apparently cheaper) possibility of modifying \eqref{action} in such a way that the problem is not there to begin with, and black holes just do not evaporate.

\subsection{An Ultra Strong Equivalence Principle for the low energy crisis}\label{sec:3.3}

The main motivation for the Ultra Strong Equivalence Principle are {\bf Problems 1} and {\bf 2} described in Sec. \ref{sec:1}. In which sense are those problems  related? Both difficulties can be associated with  the presence of spacetime dependent terms in the vacuum expectation value (VEV) of the (bare) energy momentum tensor.

\subsubsection{Statement and motivations}

Consider the vacuum expectation value of the local energy density of a (massless) field according to standard semi-classical gravity; this can be expanded at  high momenta as 
\begin{eqnarray} \label{structure2}
\langle T_0^0(x,t)\rangle_{\rm bare} &=& \int d^3 k\left(k  + \frac{f_{\rm quad}(t)}{k} + \frac{f_{\rm log}(t)}{k^3} + \dots  \right) \\[2mm]
& = &\ \  {\rm local\ terms}\ \ +\ \ {\rm non \ local\ terms}. \nonumber
\end{eqnarray}
Here spatial homogeneity has been assumed for simplicity and the $f$s are functions of time of appropriate dimensions (see eq. \eqref{square} below, or reference \cite{fulling2}, for the explicit expression in the case of a massive scalar in a flat FRW Universe). The stress tensor renormalization is a well-defined prescription to make sense of the infinities appearing in \eqref{structure2}: the local terms can be subtracted by local gravitational counterterms (cosmological constant, Newton constant and higher order local operators). The finite non-local pieces left represent the genuine particle/energy content of the choosen ``vacuum" state. 

It is interesting to note that some of the difficulties afflicting the standard low-energy framework for gravity seem to be related to the spacetime dependent subleading terms (the $f$s in the case considered above) in the energy momentum VEV expansion. The black hole information loss paradox is related to the presence of the non-local pieces, as they are the ones responsible for particle creation and for the evaporation of the black hole. We argue that also the cosmological constant problem is related to the presence of those terms, at least if we restrict the analysis to the level of non-quantized semi-classical gravity. Such a problem appears in the renormalization procedure that we described earlier as the huge amount of fine tuning that is required to match the first quartically divergent piece of \eqref{structure2} with its observed value. However, the quartically divergent term is not particularly worrisome by itself, essentially because it is a constant. In flat space we deal with it, by normal ordering or just by subtracting a constant, albeit infinite, contribution. In other words, as brutal as it might appear, ``the vacuum does not gravitate" is a well defined prescription, as long as the contribution to be neglected is a spacetime-independent scalar. We are arguing that 
the main source of trouble is not the first term in \eqref{structure2}, but  the remaining spacetime dependent pieces, as they make the normal ordering prescription meaningless and the renormalization procedure necessary in the first place.

Based on the above reasoning, we conjecture that 
\emph{there exists a theory that resembles standard semiclassical gravity at small scales but in which, as in flat space, all time-dependent pieces of \eqref{structure} (the $f$s) at any point-event, and for some ``vacuum" state, just do not exist}. 
Since the terms that we want to get rid of are effectively IR with respect to the first quartically divergent part and can be calculated, we should be able to probe this theory by IR-completing standard semi-classical gravity.

\begin{quote}
{\bf Ultra Strong Equivalence Principle (USEP):}
For each matter field or sector sufficiently decoupled from all other matter fields, there exists a state, the ``vacuum", for which the expectation value of the (bare) energy momentum tensor reads the same as in flat space, regardless of the configuration of the gravitational field.
\end{quote}

\subsubsection{Extensive quantities get holographic corrections}

In order to see more precisely how the standard geometrical description can be modified in the IR I recall briefly the analysis of by Sec. \ref{sec:2.2}. There, I showed that the correspondence \emph{submanifold-subsystem} is encoded in $(i)$ the choice of  local fields  $A(x^i, t)$ and $(ii)$ the prescription to associate to the region of space $P$ the operators of the matter field theory defined by~\eqref{bah}:
\begin{equation} \label{bah2}
{\cal A}(P,t) = \int_P d^3 x \sqrt{-g^{(3)}}\ A(x^i, t) . \qquad {\rm (standard\ framework)}
\end{equation}
The above are still operators acting on the total Hilbert space $\cH =\cH_P \otimes \cH_R $: non-trivially only on the quantum subsystem $\cH_P$ corresponding to the region of space that has been integrated over and as the identity on the rest of the system $\cH_{\rm R}$.

The metric-manifold description might break-down in that the prescription \eqref{bah2} to express average operators could be valid only in the flat-space/small distance limit. Regions of space, still perfectly defined as quantum subsystems, may not have the nice property that the corresponding operators integrate as in \eqref{bah}. I am going to argue that \eqref{bah2} should receive corrections proportional to the curvature and that it valid only up to order ${\cal O}(R V^{2/3})$, where $R$ is some curvature scalar and $V$ the volume of the region $P$: 
\begin{equation} \label{2}
A(P, t) \ \simeq \  \int_V d^3 x \sqrt{-g^{(3)}}\ A(x^i, t) \ \left[1 + {\cal O}(R V^{2/3})\right] . \qquad {\rm (modified\ framework)}
\end{equation}
I am deliberately mimicking the general type of corrections \eqref{1} that non-extensive geometrical quantities undergo in the transition from flat space to curved space. The idea is to extend this type of behavior also to extensive quantities, that in the standard description are just proportional to the volume. 

The correction in \eqref{2} is of the holographic type, i.e., proportional to the area.
It would be interesting to explore possible connections between the present approach and holography~\cite{hol1,hol2,hol3,hol4},  the idea according to which the number of degrees of freedom that a region of space can possibly contain scales as the area of the region itself. By enforcing holography on the metric manifold of semiclassical gravity one imposes a decimation of the degrees of freedom, that, in a local theory, would naturally scale as the volume. Here, I am proposing that the local description~\eqref{action} is strictly valid, but only in the small distance limit. By changing the geometrical description on the largest scales, USEP aims for a description that, in the vicinity of each point event, is possibly ``ultra local", in that excludes the non local gravitational effects that contribute to the energy momentum tensor.

The implications of \eqref{2} could be mistaken for non-local effects. Consider, for instance, a
perfectly homogeneous Universe. By definition, each comoving observer  measures in its surrounding the same energy density $\rho$. Eq. \eqref{2} implies that in that Universe, if one starts considering regions of space of Hubble size, the total energy 
inside that region will drastically differ from the three-dimensional integral of the local densities measured by the observers living therein.
In a metric manifold, that would be a dramatically non-local effect. 

\subsection{Making the energy momentum tensor look like in flat space}

In this section I explore the conjectured theory by expanding around GR. I
first review the calculation of the vacuum energy density in standard semiclassical gravity which is the ``zeroth order" approximation. Then I find the correction to the Fourier mode operators that kill the quadratic, time dependent divergence.

\subsubsection{The Zeroth Order: General Relativity, a Flat FRW Universe} \label{sec:3.4.1}

Consider a spectator free scalar field
\begin{equation} \label{action2}
S = \frac{1}{2}\int d^4 x \sqrt{-g}\left(\partial \phi^2 - m^2 \phi^2\right)
\end{equation}
in a spatially flat FRW, with metric element 
\begin{equation}
d s^2 = dt^2 - a^2(t)(d \vec{x}^2) = a^2(\tau) (d\tau^2 - d \vec{x}^2).
\end{equation}
From the above two ingredients the equation of motions of the field, 
\begin{equation} \label{equation}
\ddot \phi(t, \vec{x}) + 3 H \dot \phi (t, \vec{x}) - \frac{\partial_i^2}{a(t)^2}  \phi (t, \vec{x}) + m^2 \phi (t, \vec{x}) = 0,
\end{equation}
 and its energy momentum tensor, $T^{\mu \nu} = \frac{2}{\sqrt{-g}}\frac{\delta S}{\delta g^{\mu \nu}}$ are derivable. From that we get the expression of the energy density of $\phi$ at time $t$ and comoving spatial coordinate $\vec{x}$:
\begin{equation} \label{calH}
T^0_0(t, \vec{x}) = {\cal H}(t, \vec{x}) = \frac{1}{2}\left[{\dot \phi}^2 + \frac{1}{a^2} (\partial_j \phi)^2 + m^2 \phi^2\right].
\end{equation}
A dot means derivation with respect to the proper time $t$ while a prime denotes derivation 
with respect to conformal time $\tau$; $H = {\dot a}/a$ is the Hubble parameter. Upon quantization in the Heisenberg picture, the field can be expanded in creators and annihilators:
\begin{equation} \label{field}
\phi(t, \vec{x}) = \frac{1}{(2 \pi)^{3/2}}\int d n^3\left[\psi_n(t) e^{i \vec{n} \cdot \vec{x}} A^{(0)}_{\vec{n}} + \psi^*_n(t) e^{-i \vec{n} \cdot \vec{x}} A^{(0) \dagger}_{\vec{n}}\right].
\end{equation}
In the above equation $\vec{n}$ the conserved comoving momentum label, related to the proper physical momentum $\vec{p}$ by $\vec{p} = \vec{n}/a(t)$. The suffix
$(0)$ stands for ``zeroth order" since the creation operators just introduced are intended to be the zeroth order (--metric manifold) approximation of the operators of the IR-completed theory. 

From the action \eqref{action2} one derives the conjugate momentum $ \pi(t, \vec{x}) = \delta S/\delta \phi (t, \vec{x}),$ and from the canonical commutation relations  
$[\phi (\vec{x}), \pi(\vec{x}\, ')] = i\delta^3(\vec{x}-\vec{x}\, '),$
one derives the canonical commutators for the zeroth order-- global operators
\begin{equation} \label{normalcom}
[A^{(0)}_{\vec{n}}, A^{(0) \dagger}_{\vec{n} '}] = \delta^3(\vec{n} - \vec{n}\, ')\, .
\end{equation}
It is customary to choose $A^{(0)}_{\vec{n}}$ as the operator that always annihilate the vacuum. The chosen vacuum state is therefore implicitely characterized  by the choice of the 
mode functions $\psi_n(t)$, that, by \eqref{field} and \eqref{equation}, satisfy
\begin{equation} \label{modes0}
\ddot \psi_n + 3 H \dot \psi_n +  \omega_n^2 \psi_n = 0 ,
\end{equation}
where
\begin{equation}
\omega_n = \sqrt{\frac{n^2}{a^2} + m^2} = \sqrt{p^2 + m^2}
\end{equation}

The boundary conditions for the mode functions $\psi_n(t)$ fix the choice of vacuum.
 The argument in favor of the \emph{adiabatic vacuum} \cite{birrell,fulling,fulling2} is its resemblance to the flat space behavior for those modes that are deep inside the Hubble scale.  The adiabatic vacuum corresponds to the choice
\begin{equation} \label{adiabatic}
\psi_n(t) = \frac{1}{\sqrt{2 a^3 W_n}} \exp \left(-i \int^t W_n dt'\right),
\end{equation}
where the function $W_n(t)$ can be approximated at arbitrary adiabatic order with a formal WKB expansion. I am interested only in the next to leading order in the momentum $n$ expansion. For this purpose the approximate expression for $W_n$ is simply~\cite{fulling,fulling2}
\begin{equation} \label{adiabatic2}
W_n(t) = \omega_n(t).
\end{equation}

The vacuum expectation value of the energy density at a given point is calculated straightforwardly:
\begin{eqnarray} \label{bahbah}
\langle 0 |\, T_0^0 (t, \vec{x}) \, | 0 \rangle & = & \frac{1}{2(2\pi)^3}
\int d^3 n \, d^3 n' \left[{\dot \psi}_n {\dot \psi}^*_{n'} +\left(m^2 -  \frac{\vec{n}\cdot \vec{n}'}{a^2}\right) \psi_n \psi^*_{n'}\right][A_n^{(0)}, A_{- n'}^{(0)\dagger}]\\
& = & \frac{1}{4 \pi^2} \int_0^\infty n^2 d n \, (|{\dot \psi}_n|^2 + \omega_n^2 |\psi_n|^2),
\end{eqnarray}
where commutation relations \eqref{normalcom} have been used in the second line. From \eqref{adiabatic} and then \eqref{adiabatic2} we get
\begin{eqnarray} 
(|{\dot \psi}_n|^2 + \omega_n^2 |\psi_n|^2) &=& \frac{1}{2 a^3 W_n}\left[\omega_n^2 + W_n^2 +\frac{1}{4} ( \ln a^3 W_n){\dot \ } ^2\right]\\
& = & \frac{1}{a^3 \omega_n}\left[\omega_n^2 + \frac{1}{8} \left(3 H + \frac{{\dot \omega}_n}{\omega_n}\right)^2 + {\cal O} (n^{-2})\right],\nonumber
\end{eqnarray}
that finally gives
\begin{equation} \label{square}
\langle 0 |\, T_0^0 (t, \vec{x}) \, | 0 \rangle = \frac{1}{4 \pi^2 a^3} 
\int_0^\infty n^2 d n \, \left[\omega_n + \frac{H^2 a }{2 n} + {\cal O}(n^{-3})\right].
\end{equation}
Such a result should be opposed to the flat space result
\begin{equation} \label{square2}
\langle 0 |\, T_0^0 (t, \vec{x}) \, | 0 \rangle_{\rm flat} = \frac{1}{4 \pi^2 a^3} 
\int_0^\infty n^2 d n \, \omega_n .
\end{equation}
In order to enforce the USEP we need to find a way to get rid of the term $H^2 a/2 n$ with a IR modification.

\subsubsection{Getting rid of Quadratic Divergences} \label{sec4.5}

In order to enforce the Ultra Strong Equivalence Principle we want to get rid of the terms in the local energy momentum VEV that we normally would not find in flat space. In this paper we deal  only with the time dependent quadratic divergence, the second term inside the square brackets in \eqref{square}.

In order to fix the idea I will first consider a compact universe of total size $2 \pi a(t) L \ll H^{-1}$ much smaller than its own Hubble radius. The reason why I am doing this is that, according to our assumptions, as long as that condition is satisfied, the departure from usual semi-classical gravity are extremely small: this space-time, even globally, can be described as a manifold to a very good approximation. 
Such an hypothesis is not mandatory; on the opposite, it will turn (see section~\ref{sec:3.4.3}) out that a spatially compact Universe is incompatible with the proposed IR modification. Nevertheless, a finite size Universe much smaller than its Hubble radius helps fixing the ideas and finding the correct recipe. 

First of all, I focus our attention on a given comoving observer/trajectory at any proper time $t$ and, say, at point $\vec{x} = 0$. For brevity, when we write $\vec{x} \approx 0$, we refer to a region of space around that trajectory small enough that the manifold description holds and we can define spatial gradients and local commutation relations between neighboring points. 
I will not touch the local QFT quantities but we will allow the globally defined operators (e.g. the ``Fourier modes") to  get ${\cal O}(L H)^2$ corrections. 
So I start by postulating the local equations that are untouched, the
field equations \eqref{equation} in the Heisenberg picture for the field operator in $\vec{x} = 0$,
\begin{equation} \label{evo}
\ddot \phi(t, \vec{x} \approx 0) + 3 H \dot \phi (t, \vec{x} \approx 0) + \left(m^2 - \frac{\partial_i^2}{a^2}\right)  \phi (t, \vec{x} \approx 0)= 0
\end{equation}
and the Hamiltonian density ${\cal H}(t, \vec{x}\approx 0)$ at point 0, which, again, is an explicitely time dependent operator:
\begin{equation} \label{hamtime}
T^0_0 =  {\cal H}(t, \vec{x} \approx 0) = \frac{1}{2}\left[ \dot{\phi}^2 (t, \vec{x} \approx 0) + \frac{\partial_i \phi^2 (t, \vec{x} \approx 0)}{a^2} + m^2 \phi^2 (t, \vec{x} \approx 0) \right] .
\end{equation}
While \eqref{equation} and \eqref{calH} are both elegantly derivable from the same action, for the moment we are forced to consider \eqref{evo} and \eqref{hamtime} as two basic -- and in principle independent -- ingredients of this novel framework. Note moreover that, in a sort of pre-geometric fashion,
we can take the above equations as the implicit definition of the local expansion $a(t)$. Commutation relations between local quantities are strictly preserved:
\begin{equation} \label{local2}
[\phi (\vec{x}), \pi(\vec{x}' \approx \vec{x})] = i\delta^3(\vec{x}-\vec{x}'),
\end{equation}
(the momentum conjugate to $\phi$ can still be defined completely locally in terms of the Lagrangian density in $\vec{x} = 0$: $\pi(\vec{x}\approx 0) = \partial {\cal L}(\vec{x}\approx 0)/\partial \dot \phi (\vec{x}\approx 0)$).

By definition, the Fourier modes  $\phi_{\vec{n}}$ of the field satisfy (I now use a finite volume version of \eqref{field})
\begin{equation} \label{fourier}
\phi(t, \vec{x}\approx 0) = \frac{1}{(2 \pi L)^3} \sum_{\vec{n}} \phi_{\vec{n}} (t) \ e^{i \vec{n} \cdot \vec{x}},
\end{equation}
where 
\begin{equation} \label{phik}
\phi_{\vec{n}} = \psi_n(t)  A^{(1)}_{\vec{n}} + \psi^*_n(t)  A^{(1)\, \dagger}_{\vec{n}} .
\end{equation}
The index $(1)$ means that I am now dealing with the corrected operators.
Again, at the exponent on the RHS the coordinates should be considered to extend as far as we need to define the derivatives of the field. Note that the coordinate $\vec{x}$ and momenta $\vec{n}$ are comoving with respect to the local expansion $a(t)$: $\vec{x} = \vec{x}_{\rm phys} / a(t)$, $\vec{n} = a(t) \vec{n}_{\rm phys}$ .
Since we are in a compact space, we expect  ${\vec{n}}$ to take discrete values and,  at least at zeroth order in our approximation, we have $n_i \simeq N_i/L$, the components $N_i$ being integer entries. By \eqref{fourier}, ${\vec{n}}$ is still the derivative at 0 in Fourier space: 
$\partial_i \ = i n_i$.

The commutator between ${ A}$ and ${ A}^\dagger$ is proportional to the total volume
and therefore will receive the postulated corrections \eqref{2}. We make the following ansatz:
\begin{equation}  \label{piazza}
[{ A}^{(1)}_{\vec{n}},{ A}^{(1)\, \dagger}_{\vec{n}'}] = (2 \pi L)^3 \ \delta_{{\vec{n}}, {\vec{n}'}}\ \left(
1 - \gamma \frac{H^2}{n_{\rm phys}^2} + \dots \right) \, .
\end{equation}
Since $n>1/L$ (the zero mode will not be considered), the correction is of the type conjectured in \eqref{2}, i.e., ${\cal O} (L^2 R)$. The parameter $\gamma$ will be determined by applying the Ultra-Strong equivalence principle to the quadratically divergent part of the stress-energy tensor VEV.

 The corrected Fourier modes are effectively time dependent. At least at this order of approximation, however, the time dependence is just a multiplicative factor that leaves the vacuum untouched. Note also that, by applying the postulated local equation \eqref{evo}, the time dependence of $A^{(1)}_n$ introduces additional spurious terms inside \eqref{modes0}. Those, however, are of higher order in the $n$ expansion and therefore do not affect the present result. Finally, inverse Fourier transforms  are not allowed. Following the reasoning of section \ref{sec3}, we argue that whenever we take integrals $\int d^3 x$ of operators over a volume $V$ we are making a mistake of order ${\cal O}(H^2 V^{2/3})$.

Up to the order of the quadratic divergence, the calculation of the energy momentum tensor VEV proceeds as before, with the only difference that, inside \eqref{bahbah}, Fourier integrals are substituted by series and $[A_n^{(0)}, A_{- n'}^{(0)\dagger}]$ by \eqref{piazza}. This gives
\begin{equation} \label{orderbyorder}
\langle 0 |\, T_0^0 (0) \, | 0 \rangle \sim
\sum_{\vec{n}} \left(\frac{n}{a}\right) \left(1 + \frac{H^2 a^2}{2 n^2} +
{\cal O}(n)^{-4} \right)\left(1 - \gamma \frac{H^2 a^2}{n ^2}\right)
\end{equation}
where the last factor comes from the anomalous  commutation relations \eqref{piazza}
Note that the quadratic divergence can always be reabsorbed by setting $\gamma = 1/2$. This gives the modified commutation relation
\begin{equation}
[{ A}^{(1)}_{\vec{n}},{ A}^{(1)\, \dagger}_{\vec{n}'}] = (2 \pi L)^3 \ \delta_{{\vec{n}}, {\vec{n}'}}\ \left(
1 -  \frac{H^2 a^2}{2 n^2}  \right) \, .
\end{equation}

\subsubsection{USEP is not supported in a compact universe}\label{sec:3.4.3}

Although I previously referred in the text to modifications at scales of the inverse curvature, it is the extrinsic curvature $\sim H$ which appears to regulate the correct modification. Equivalently, we can just say that the annihilators are modified with respect to GR by
\begin{equation} \label{squareroot}
A_{\vec{n}}^{(1)} = \sqrt{1 - \frac{H^2 a^2}{2 n^2}} A_{\vec{n}}^{(0)} .
\end{equation}

So far we concentrated only on local operators at some given point $\vec{x}=0$ and made connection with the global Fourier modes. A crucial ingredient to understand the large-scales 
properties of this deformed spacetime is the momentum operator constructed with the modified Fourier modes,
\begin{equation} 
{\vec P}  =  \frac{1}{(2 \pi L)^3} \sum_{\vec{n}} \vec{n}\,  A^{(1) \dagger}_{\vec{n}} A^{(1)}_{\vec{n}}\, .
\end{equation}
Equivalently, by using \eqref{squareroot} the latter can be written as 
\begin{equation} \label{momentum1}
{\vec P}  = \frac{1}{(2 \pi L)^3} \sum_{\vec{n}} \vec{k}(\vec{n})\,  A^{(0) \dagger}_{\vec{n}} A^{(0)}_{\vec{n}}\, ,
\end{equation}
i.e. by using the standard GR operators $A^{(0)}$ satisfying the usual commutation relations, and 
\begin{equation}\label{kkk}
\vec{k}(\vec{n}) = \vec{n} \left(1 - \frac{H^2 a^2}{2 n^2}\right).
\end{equation}
is a modified Fourier label. We want to be able to move in this space arbitrarily far in any direction and define local operators also elsewhere. For this purpose a translation operator is need. Since homogeneity is assumed here, it is possible to obtain a global translation operator by exponentiation of $P_i$,
\begin{equation}T_i(\lambda) = e^{-i \lambda P_i} . \label{translation}
\end{equation}
Thus, we can define local operators at a finite distance from $\vec{x}\approx 0$ by the action of $T_i$:
\begin{equation} \label{keepgoing}
\phi(t, *) \equiv
T_i(\lambda)\, \phi(t, 0) \, T_i^{-1}(\lambda) =  \frac{1}{(2 \pi L)^3} \sum_{\vec{n}} \, \phi_{\vec{n}} \ e^{- i\, \lambda \, n_i \left(1- \frac{H^2 a^2}{2 n^2}\right)}.
\end{equation}
where $*$ stands for ``at distance $\lambda$ if one keeps going in the $i$ direction" and $\lambda$ is made up by all the infinitesimal steps that have been taken in order to go from one point to another and is therefore the (comoving) proper distance. 

Now remember that in this compact universe of comoving size $2 \pi L$ the $n_i$ are quantized, $n_i = N_i/L$. Normally, the exponentials of (\eqref{keepgoing}) would just be $e^{- i \lambda n_i}$ and they would all be equal to unity for $\lambda = 2 \pi L$. This means that, by going $2 \pi L$ far along one of the three directions $i$, you just take a round trip and are back to the original field quantity $\phi(0)$. However, the present modified framework looks incompatible with a compact Universe since, due to the presence of the correction terms at the exponent in \eqref{keepgoing}, for no value of $\lambda$ can you just take a round trip and obtain back the same operator $\phi(0)$. It looks that a definite prediction of the ultra strong equivalence principle is that the universe has to be infinitely extended in any direction.

\subsection{A look at the global picture}

Since the universe has to be infinitely extended in any direction we go back to Fourier integrals instead of Fourier series. We can still use the exponential of the momentum operator \eqref{momentum1},
\begin{equation} \label{momentum2}
\vec{P}^{(1)} = \int d^3 n\, \vec{k} \, A_{\vec{n}}^{(0)\dagger} A_{\vec{n}}^{(0)} ,
\end{equation}
 to define local quantities away from the origin. Again, at this order of approximation, ${\vec k}$ is given by
\begin{equation} \label{kk2}
\vec{k} = \vec{n} \left(1 - \frac{H^2 a^2}{2 n^2} \right). 
\end{equation}

The form of \eqref{momentum2}, moreover, suggests another way to look at the proposed deformation. That of  referring to the usual GR operators $A_{\vec{n}}^{(0)}$, satisfying the usual commutation relations, and modifying the momentum labels as in \eqref{kk2} instead. This approach is summarized in the following section.

\subsubsection{The recipe: a modified dispersion relation}

I am trading modified Fourier operators for modified dispersion 
relations. In other words, in this equivalent (at this order of approximation), but more practical, formulation, I am postulating a slight mismatch between the ``metric-manifold" labels $\vec{n}$ and the physical momenta $\vec{k}$ that locally define the infinitesimal translations and the derivatives of the local fields.
Since $\vec{n}$ is the usual comoving ``manifold" label, it is assumed to be time-independent. Moreover, since only the usual creators operators $A_{\vec{n}}^{(0)}$ will be used from now on, I will drop the  the suffix $(0)$ to simplify the notation. The recipe can finally be summarized as follows:

\begin{itemize}
\item There are two different (comoving) Fourier coordinates: the ``manifold"-coordinate $\vec{n}$ and the locally defined physical momentum $\vec{k}$. The integration measure in Fourier space is flat in the $n$-coordinates. Moreover, during time evolution, $\vec{n} $ is conserved, $\vec{k}$ is not. The relation between the two is
\begin{equation} \label{kk}
\vec{k} = \vec{n} \left(1 - \frac{H^2 a^2}{2 n^2} + {\rm higher\ order} \right).
\end{equation}

\item The local field in $\vec{x}\approx 0$ is expanded in Fourier modes as follows:
\begin{equation} \label{localfield}
\phi(t, \vec{x}\approx 0) = \frac{1}{(2 \pi)^{3/2}}\int d n^3\left[\psi_k(t) e^{i \vec{k} \cdot \vec{x}} A_{\vec{n}} + \psi^*_k(t) e^{-i \vec{k} \cdot \vec{x}} A^\dagger_{\vec{n}}\right] \equiv \frac{1}{(2 \pi)^{3/2}}\int d n^3 \phi_{\vec k}(t) e^{i \vec{k} \cdot \vec{x}} 
\end{equation}
and the $A_n$ satisfy the usual commutation relations: 
\begin{equation} \label{usualcom}
[A_{\vec{n}}, A^\dagger_{\vec{n}'}] = \delta^3(\vec{n} -\vec{n}').
\end{equation}

Again, the $\vec{x}$ coordinate in \eqref{localfield} extends only as far as is needed to take the space derivatives of the field in $\vec{x}=0$. When a derivative is taken, a factor of $\vec{k}$, instead of $\vec{n}$, drops. 

\item The local equation for $\phi(t,\vec{x}\approx 0)$, \eqref{evo}, fixes the equations for the functions $\psi_n$:
\begin{equation}
\ddot{\psi}_k(t)  + 3 H \dot{\psi}_k(t) +  \omega^2_k \psi_k(t) = 0, 
\end{equation}
where
\begin{equation}
\omega_k = \sqrt{\frac{k^2}{a^2} + m^2} 
\end{equation}
and $\vec{k}$ is given in \eqref{kk}. The solution of the above equations are choosen to correspond to the adiabatic vacuum (in the physical momentum $k$), i.e.
\begin{equation} \label{adiabatic3} 
\psi_k(t) = \frac{1}{\sqrt{2 a^3 W_k}} \exp \left(-i \int^t W_k dt'\right).
\end{equation}
The function $W_k$ can be approximated at arbitrary adiabatic order as before. The only difference is that $\omega_k$ should substitute $\omega_n$ in \eqref{adiabatic2}.

\item The energy density in $ \vec{x} \approx 0$ is expressed in terms of the local field $\phi(t,\vec{x}\approx 0)$ in the usual way (equation \eqref{hamtime}), 
\begin{equation} \label{energymomentum}
T^0_0 =  {\cal H}(t, \vec{x} \approx 0) = \frac{1}{2}\left[ \dot{\phi}^2 (t, \vec{x} \approx 0) + \frac{\partial_i \phi^2 (t, \vec{x} \approx 0) }{a^2} + m^2 \phi^2 (t, \vec{x} \approx 0) \right] .
\end{equation}

\end{itemize}

The above equations can be used to calculate the vacuum expectation value of $T_0^0$. By direct substitution of \eqref{localfield} into \eqref{energymomentum} we get
\begin{eqnarray} \nonumber
\langle 0 |\, T_0^0 (t, \vec{x}\approx 0) \, | 0 \rangle & = & \frac{1}{2(2\pi)^3}
\int d^3 n \, d^3 n' \left[{\dot \psi}_k {\dot \psi}^*_{k'} +\left(m^2 -  \frac{\vec{k}\cdot \vec{k}'}{a^2}\right) \psi_k \psi^*_{k'}\right][A_n, A_{- n'}^{\dagger}]\\
& = & \frac{1}{4 \pi^2} \int_0^\infty n^2 d n \, (|{\dot \psi}_k|^2 + \omega_k^2 |\psi_k|^2),
\end{eqnarray}
By substitution of \eqref{adiabatic3} we then get
\begin{eqnarray} 
\langle 0 |\, T_0^0 (t, \vec{x}\approx 0) \, | 0 \rangle &=& \frac{1}{4 \pi^2 a^3} 
\int_0^\infty n^2 d n \, \left[\omega_k + \frac{H^2 a }{2 k} + {\cal O}(n^{-3})\right], \\
& = & \frac{1}{4 \pi^2 a^3} 
\int_0^\infty n^2 d n \, \left[\omega_n  + {\cal O}(n^{-3})\right]
\end{eqnarray}

Note that the quadratic divergences are here reabsorbed just be re-expressing $\langle 0 |\, T_0^0  | 0 \rangle$ in terms of the appropriate ``flat measure" time-independent Fourier coordinates $n$.

\subsubsection{Correction to the comoving distance} \label{sec:3.5.2}

As in section \ref{sec:3.4.3}, by exponentiating the momentum operator, it is possible to define local fields away from the origin:

\begin{equation} \label{keepgoing2}
\phi(t, *) \equiv
T_i(\lambda)\, \phi(t, 0) \, T_i^{-1}(\lambda) =  \frac{1}{(2 \pi)^{3/2}} 
 \int d^3 n\, \phi_{\vec{k}} \ e^{- i\, \lambda \, n_i \left(1- \frac{H^2 a^2}{2 n^2}\right)}.
\end{equation}
where again $*$ stands for ``at (comoving) distance $\lambda$ if one keeps going in the $i$ direction" and $\lambda$ is made up by all the infinitesimal steps that have been taken in order to go from one point to another and is therefore the (comoving) proper distance.

In the present case, because of homogeneity, 
translations in different directions commute with each other, so we can label the local operators with three parameters and describe, approximately, a finite patch of Universe around $\vec{x}\approx 0$. Explicitely,

\begin{equation} \label{localfar}
\phi(t, \vec{\lambda}) =  \frac{1}{(2 \pi)^{3/2}} 
 \int d^3n\,  \left[\psi_k(t) e^{i \vec{k} \cdot \vec{\lambda}} A_{\vec{n}} + \psi^*_k(t) e^{-i \vec{k} \cdot \vec{\lambda}} A^\dagger_{\vec{n}}\right].
\end{equation}
Although the above equation has a familiar form, the labels $\lambda^i$ should really be thought here as the parameters of the translation group. In the following it is  shown that, if $\lambda^i$ are used  as integration variables at $t =$ const., as we would do on a standard FRW flat space, an error of order $V^{2/3} H^2$ is made.

It is interesting to calculate, at some given time $t$, the canonical commutator between $\pi(0) = a^3 \dot \phi(0)$ and the field itself at a distance. By using the standard commutation relations 
\eqref{usualcom} we obtain
\begin{equation}
[\pi(0), \phi( \vec{\lambda})] = - \frac{a^3}{(2\pi)^3} \int d^3 n \, \left[\dot{\psi}_k^* \psi_k e^{i \vec{k} \cdot \vec{\lambda}} - \dot{\psi}_k \psi_k^* e^{- i \vec{k} \cdot \vec{\lambda}}\right] .
\end{equation}
By using the Wronskian condition $a^3  (\dot{\psi}_k^* \psi_k - \dot{\psi}_k \psi_k^*) = i $, immediately derivable from \eqref{adiabatic3},  and \eqref{kkk} we get, up to higher orders in $H a \lambda$,
\begin{equation}
[\pi(0), \phi( \vec{\lambda})] = -\frac{i}{(2\pi)^3} \int d^3 n\, e^{- i\, \vec{\lambda}\cdot \vec{n} \left(1- \frac{H^2 a^2}{2 n^2}\right)} \simeq 
-\frac{i}{(2\pi)^3} \int d^3 n\, e^{- i\, \vec{\lambda}\cdot \vec{n}} \left(1+ i \vec{\lambda}\cdot \vec{n} \frac{H^2 a^2}{2 n^2}\right),
\end{equation} 
where we have expanded the exponential in the last expression. The first term gives the usual delta function. The last term gives
\begin{equation}
\frac{i H^2 a^2}{2 (2\pi)^3} \frac{d}{d\alpha}\left.\int \frac{d^3 n}{n^2} e^{- i\, \alpha\, \vec{\lambda}\cdot \vec{n}} \right|_{\alpha = 1}\ =\ - \frac{i}{8 \pi} \frac{H^2 a^2}{\lambda}\, .
\end{equation}
In this way we obtain the first correction to the canonical commutator, 
\begin{equation} \label{comm}
[\pi(0), \phi( \vec{\lambda})] = -i\left(\delta^3(\vec{\lambda}) + \frac{1}{8 \pi} \frac{H^2 a^2}{\lambda} \right).
\end{equation}
The above relation finally gives a more precise sense to \eqref{local2}. At face value, \eqref{comm} is a signal of a non-locality that becomes more and more important at large scales. But again, that is the interpretation that we would have of \eqref{comm} on a metric manifold. 
In section \ref{sec3} we argued that regions of space at a given time can still be associated with a well defined algebra of ``averaged" operators of the matter theory, but that those operators can only approximately be expressed as integrals over a metric manifold. The expression in \eqref{comm} is in fact a three-dimensional density and the correction term is just signaling the mistake that we make if we naively take the integral 
\begin{equation}
\int_{\lambda a\ll H^{-1}}  d^3 \lambda'\, [\pi(0), \phi(\vec{\lambda}')] \ =\  -i \left[1 + \frac{1}{4} H^2 a^2 \lambda^2 +{\cal O}(H^2 V^{2/3})^2 \right]
\end{equation}
and use $\lambda$ as a flat-measure integration variable.
We nicely recover the expected pattern \eqref{2}.

It is equally interesting to calculate $[\phi(0), \pi(\lambda)]$. First, we note that there is a potential ambiguity in defining the time derivative of a displaced operator. 
By deriving \eqref{localfar} we get
\begin{eqnarray} 
 \pi(\vec{\lambda}) &=&  \frac{a^3}{(2 \pi)^{3/2}} \nonumber
 \int d^3n\,  \left[\dot{\psi}_k(t) e^{i \vec{k} \cdot \vec{\lambda}} A_{\vec{n}} + \dot{\psi}^*_k(t) e^{-i \vec{k} \cdot \vec{\lambda}} A^\dagger_{\vec{n}}\right] \\
 &+& i \frac{a^3}{(2 \pi)^{3/2}}  \int d^3n\,  (\vec{k} \cdot \vec{\lambda})\dot{}\ \left[\psi_k  \ e^{i \vec{k} \cdot \vec{\lambda}} A_{\vec{n}}  - \psi^*_k  \ e^{- i \vec{k} \cdot \vec{\lambda}} A^\dagger_{\vec{n}} \right] . \label{ambig}
\end{eqnarray}
The second line in the above equation is there because $k$ is time dependent. 
However, if we just applied the translation to $\pi(0)$, instead of deriving $\phi(\vec{\lambda})$, those terms would not be there. Therefore, for consistency, we need to make them ineffective at the required order of approximation. In other words, we have to impose that  $[\phi(0), \pi(\vec{\lambda})] = - [\pi(0),\phi(\vec{\lambda})]$. Because of the second line of \eqref{ambig}, the commutator between $\phi(0)$ and $\pi(\lambda)$ gives
\begin{equation} \label{nonsym}
[\phi(0), \pi(\vec{\lambda})] = - [\pi(0),\phi(\vec{\lambda})] - 2i \frac{a^3}{(2\pi)^3}
\int d^3 n \, e^{- i \vec{k} \cdot \vec{\lambda}} \left| \psi_n \right|^2 (\vec{k} \cdot \vec{\lambda})\dot{} \ ,
\end{equation}
the last term being the spurious contribution. In order to get rid of it, we have to  make the comoving physical distance $\vec{\lambda}$ also time dependent. This effectively means that, after an infinitesimal time step $dt$, we have to reconsider the field translated, from  $\vec{x} \approx 0$, not by the same comoving distance $\lambda$, but by a slightly different amount. 
At high momenta/small distances, since $|\psi_k|^2 \sim 1/n$, the integral in the second term of \eqref{nonsym} reads
\begin{equation}
\int d^3 n \, e^{- i \vec{n}  \cdot \vec{\lambda}} \, \frac{1}{n} \left[\dot{\vec{\lambda}} \cdot \vec{n}\left(1 - \frac{H^2 a^2}{2 n^2}\right) - \vec{\lambda}\cdot \vec{n} \frac{(H^2 a^2)\dot{}}{2 n^2}\right]\, .
\end{equation}
We make the ansatz $\dot{\vec{\lambda}} = \beta \lambda^2 (H^2 a^2)\dot{}\, \vec{\lambda}$, $\beta$ being a number to be determined. We get
\begin{equation}
\int d^3 n \, e^{- i \vec{k} \cdot \vec{\lambda}} \left| \psi_n \right|^2 (\vec{k} \cdot \vec{\lambda})\dot{}\ \simeq \ i  (H^2 a^2)\dot{} \ \frac{ d}{ d \alpha} \int d^3 n\, \left. \left(\frac{\beta \lambda^2}{n} - \frac{1}{2 n^3}\right) e^{- i \alpha\, \vec{n} \cdot \vec{\lambda}}\right|_{\alpha =1}.
\end{equation}
The last integral can be regularized by setting $n^{-3} \rightarrow n^{-3 + \epsilon}$ and taking the $\epsilon \rightarrow 0$ limit only after deriving with respect to $\alpha$. The result is null for $\beta = 1/4$, which fixes the time dependence of $\lambda$:
\begin{equation} \label{expansion}
\dot{\lambda} = \lambda^3 \frac{(H^2 a^2)\dot{}}{4} .
\end{equation}

The above formula is one of the main results of this paper and can be rephrased as follows. In this homogeneous Universe, in the vicinity of each point, there is a local expansion given by the same scale factor $a(t)$. In essence, $a(t)$ rules the evolution equations for the local operators and determines the rate at which  the distance between two comoving observers grows, as far as such a distance is extremely small compared to Hubble. In that limit, the comoving distance $\lambda$ is conserved.  
However, if you pick up a pair of comoving observers further apart, their distance 
grows slightly differently, which means that the expansion rate on large scales is effectively dependent on the distance. The modified expansion rate is given, at first order, by \eqref{expansion}. This effect, which is the most striking signal of the breakdown of the manifold description, is clearly negligible well within the Hubble scale.

\subsubsection{Cosmology}

Modifications of General Relativity (GR) on the largest scales have been advocated~\cite{dgp,dgp2} in order to explain the present acceleration of the Universe. IR modifications of GR look now particularly appealing~\cite{justin3,justin4} in view of some emerging tensions between standard $\Lambda$CDM cosmology and large-scale observations. Those include the lack of large-scale CMB correlations~\cite{co1,co2,co3}, the anomalously large
integrated Sachs-Wolfe cross-correlation~\cite{isw}, the abnormally large bulk flows on very large scales~\cite{ab1,ab2} a  (marginal) mismatch between angular diameter- and luminosity- distances at high redshift \cite{bruce,will} and the anomalous growth of large scale structures from weak lensing measurements~\cite{rachel}. 

The IR modifications of GR considered so far are generally equivalent to giving the graviton a small mass.  
The models of massive gravity  follow a basic pattern that is common to all dark energy models: there is a tuned scale ``hidden" in the theory which becomes effective, ``by coincidence", when the mass of the graviton $m_g \sim H_0$ drops to about its value. 
The USEP and, more general, the type of modifications that I am considering here, suggests an alternative which looks appealing. As much as GR is a geometrical deformation of flat space at distances of the order of
the curvature radius, eq. \eqref{expansion} suggests the possibility of a 
subleading curvature-dependent effect that systematically modifies further the geometrical description of GR at large distances. 
High redshift observations have the unique property
of relating objects (e.g., the observer and the supernova) that
are placed from each other at a relative distance of the order $H^{-1}$.
Therefore, modifying GR in the IR at a
length scale set by the curvature -- rather than fixed \emph{a
priori} by a parameter -- will systematically affect any
cosmological observation at high redshift, regardless of when such
an observation takes place and without the need of any external
mass scale.

Eq. \eqref{expansion} means that, on large scales, the expansion effectively detaches from the local expansion rate ${\dot a}/a$. By direct integration it is easy to show that  the proper distance $d = a \lambda$ between two comoving observers evolves from time $t'$ to $t$  as
\begin{equation} \label{d}
\frac{d(t)}{d(t')} = \frac{a(t)}{a(t')}\left[1 - \frac{d^2(t')}{2}\left(H^2(t) \frac{a^2(t)}{a^2(t')} - H^2(t')\right)\right]^{-1/2}.
\end{equation}
According to the equation above the distance between two comoving observers grows slightly less than in a FRW Universe with scale factor $a(t)$.

We now want to calculate the ``corrected" trajectory $r(\tau)$ for a light ray, where $r$ is its comoving distance and $\tau$ the conformal time. 
A light ray passing through $\vec{x}\approx 0$ satisfies $d r = d\tau$. However, on top of the usual contribution, the trajectory of a light ray receives a correction from the modified global expansion \eqref{expansion}:
\begin{equation}
\frac{d r}{d \tau} = 1 + r^3 \frac{(H^2 a^2)'}{4} .
\end{equation}
Now we specialize to a matter dominated Universe for simplicity.  Therefore, we set $H a = 2/\tau$ and we get
\begin{equation} \label{22}
\frac{d r}{d\tau} = 1 - 2\frac{r^3}{\tau^3}.
\end{equation}
Normally, at the RHS we would simply have $1$ and, again, the second term should be considered just as the first order correction.
With respect to an Einstein-De Sitter Universe, a light ray propagating forward in time will go a shorter way. A light ray that we are detecting now, on the other hand, is actually coming from farer.  For a light ray propagating back in the past the above equation becomes, in terms of the redshift,
\begin{equation} \label{backintime}
\frac{d (H_0 r)}{dz} = \frac{1}{(1+z)^{3/2}} +  \frac{(H_0 r)^3}{4}.
\end{equation}
The above should be solved with initial condition $r(0) = 0$ and will give some $r(z)$ to be used in what follows. Note that, as in the normal framework, the Hubble parameter $H_0$ is just a total normalization factor for cosmological distances. 

Light rays going backward in time in any direction describe, at some redshift $z$, the surface of a two-sphere of area $4 \pi a(z)^2 r(z)^2$. Therefore, 
the angular diameter distance is $r(z)$ multiplied by $a(z)$:
\begin{equation}
d_A(z) = \frac{1}{1+z} r(z).
\end{equation}
Angular diameter and luminosity distances are related~\cite{ircompletion} by the usual formula 
\begin{equation}
d_L = (1+z)^2 d_A\, .
\end{equation}

The correction \eqref{backintime} goes in the direction of a positive acceleration in that it increases the luminosity distance of a  matter dominated universe. However, this effect is not enough, by itself, to explain the observed amount of acceleration. In particular, the luminosity distance of a matter dominated universe is affected by \eqref{backintime} only at forth order in the redshift, while the presence of a dark energy component kicks in at second order. 

The compatibility of USEP with supernovae data has been the subject of a recent paper~\cite{savvas}. We have considered a wider class of IR modifications of GR, namely, instead of \eqref{expansion}, 
\begin{eqnarray}
\dot \lambda & = & A_1\,  \lambda H + A_2 \,  \lambda^2 H^2 a + \dots \nonumber \\
& & +B_1 \, \lambda^2 (H a)^{\cdot} + B_2 \, \lambda^3  (H^2 a^2)^{\cdot} + \dots\, .
\label{dotlam}
\end{eqnarray}

The terms in the above expansion rearrange when we calculate the luminosity
distance. Therefore, for a matter-dominated universe, a quite
general structure of subleading terms that generalizes \eqref{backintime} is given by
\begin{equation} \label{ODE}
\frac{d (H_0 r)}{d z} = \frac{1}{(1+z)^{3/2}} F\left(H_0 r (1+z)^{1/2}\right),
\end{equation}
where $F(x)$ is a generic function with $F(0) = 1$:
\begin{equation} \label{ODE2}
F(x) = 1 + \alpha x + \beta x^2 + \gamma x^3+\, \dots \, .
\end{equation}
We fitted the parameters in \eqref{ODE2} to study the compatibility of supernovae data with a class of IR modifications of gravity that do not contain any new mass parameter. Eq.~(\ref{backintime})
corresponds to $\alpha = \beta = 0$ and $\gamma=1/4$, while in GR
all coefficients are set to zero.

As a by-product, the analysis of ~\cite{savvas} shows that the model in \eqref{backintime}, i.e., the particular correction given by USEP, is 20$\sigma$ away from the $\Lambda$CDM best fit and therefore strongly disfavored by data. The reason is that the correction in ~\cite{savvas} is too weak, of too high order in the redshift, to give account, by itself, of the measured amount of acceleration. Intriguingly, we have found that the model \eqref{ODE}-\eqref{ODE2} fits
the data very well for the values close to the exact numbers
$\alpha=1$ and $\beta=-1/2$, that give the same $\chi^2$ as $\Lambda$CDM. It is interesting to consider such sharp
numerical values, not because of abstract numerology, but because
a mechanism analogous to that described in the last section
very naturally produces coefficients which are integers or simple
fractions. The suggested luminosity distance may eventually turn out to be
produced by considering analogous IR-modifications produced by apply USEP to other fields\footnote{For instance, a
different mechanism, based on a Casimir-like vacuum
energy~\cite{urban1,urban2}, needs a Veneziano ghost in order reproduce a
density of  the right order of magnitude. For yet another IR approach to dark energy see e.g.~\cite{varun}} or derived with different theoretical insights.

\subsection{Discussion}

I have tried to modify the IR-side of the paradigm ``GR + matter fields" and  look at \eqref{action} as a small distance approximation. The proposed modification is radical because it challenges the usual geometrical description of spacetime on the largest scale, but compelling because it does not contain any adjustable parameter: in a cosmological set-up deviations from GR are calculable and become relevant, at any epoch, at the Hubble scale. Abandoning the well understood framework \eqref{action} is adventurous; but I have tried to do so by following a physical principle that seems to address, at least in part, {\bf Problems 1} and {\bf 2} of Sec. \ref{sec:1} and, at the same time, to give interesting cosmological predictions at high redshift. 

This model suggests is that the present acceleration of the Universe may not be due to any additional dark energy component but to a systematic  breakdown of the geometrical description of spacetime at length scales comparable to Hubble. The proposed effect does not kick in at some given time ``by coincidence" during the cosmological evolution, but is permanently present at any cosmological epoch whenever a comoving observer looks at very far away objects.

The theoretical idea at the basis of this modification is the ultra strong equivalence principle: the bare VEV of the energy momentum tensor is the same as in flat space. It is clearly not guaranteed that a theory with such properties exists in full generality beyond the example considered here; if it does, its formulation will certainly require a major theoretical breakthrough. Giving up the metric manifold structure constitutes a significant technical difficulty. On the other hand, compared to some of the subtleties of standard QFT on curved spacetime~\cite{birrell}, the physical picture that I am considering is somewhat simplified, made trivial. I am  conjecturing the possibility of describing, locally, the physical phenomena in exactly the same way as we do in flat space, and that the large scale-properties of spacetime have no effect whatsoever inside Einstein's free-falling elevator.

The reader that doesn't find this approach particularly realistic or convincing might still be curious to answer the following question: how much should we modify GR (the global Fourier modes operators as in \eqref{squareroot} or the dispersion relations as in \eqref{kk2}) in order to ``renormalize" the stress tensor with an IR modification instead than with local gravitational counterterms? In the last pages I have shown that such a modification is in fact extremely tiny. For a free scalar field in a cosmological set-up, such a modification is weaker than the one needed to explain the current acceleration of the universe, although it goes in the same direction.

It would be interesting to consider the implications of this model in the very early Universe and close to the singularity, where IR corrections of order Hubble might become important. 
|t is also interesting to note that the usual cosmological kinematic of the modes (``exiting" during acceleration, ``entering" during deceleration) is modified in the present framework, as the physical comoving momentum is not conserved on large scales (${\vec n}$ is conserved, not ${\vec k}$, see eq. \ref{kk2}). It would be interesting to see what are the implications of this effect in the early Universe, although higher orders in the IR expansion might be needed in order to have a more complete picture.

\section{Alternative Localization Scheme} \label{sec4}

In the last section I exploited the \emph{subsystem-description} by considering the possibility of a break-down of the metric \emph{manifold-description} at large distances. Another, more conservative, direction that exploits the \emph{subsystem description} is that of considering alternative \emph{localizations}~\cite{fabio1,sergio}. The idea is to keep both the complementary descriptions \emph{submanifold/subsystem} and address the appropriate rationale to make such a correspondence. I will briefly described the standard localization scheme and then consider an alternative localization, proposed by Newton and Wigner sixty years ago~\cite{nw} with the aim of finding a well defined position operator in relativistic quantum theory.  The Newton-Wigner and the standard localizations differ at scales of order the Compton wavelength of the field or the temperature/average energy of the state considered and therefore largely agree on the usual macroscopic scales where gravity is so-well tested. I will try to explore some of the implications for semiclassical gravity derived by considering such a different fine-grained spacetime description.

\subsection{Standard and Newton-Wigner localizations} \label{sec:4.1}

In order to see how regions of space are described in standard semiclassical gravity I will now make more explicit the underlying correspondence \emph{submanifold/subsystem}. To be concrete I will consider a free scalar field $\phi$ of mass $m$ in Minkowski space. The Hamiltonian reads 
\begin{equation} \label{sec4:ham}
H\ =\ \frac{1}{2} \int d^3 x \left(\pi^2 + \nabla \phi^2 + m^2 \phi^2\right)\ =\ \int d^3 k \, w_k \, \ada_\kk\, a_\kk\, ,
\end{equation}
where the infinite vacuum contribution has been subtracted in the last equality and $w_k = \sqrt{{ k}^2 + m^2}$. The standard commutation relations between $\pi(\x) = {\dot \phi}(\x)$ and $\phi$ read
\begin{equation} \label{sec4:commu}
[\phi (\vec{x}), \pi(\vec{x}\, ')] = i\delta^3(\vec{x}-\vec{x}\, '),
\end{equation}
while operators $a_\kk$ also satisfy 
\begin{equation}
[a_\kk, a_\kp]=0,\,  [a_\kk, \ada_\kp]=\delta^3(\kk - \kp).
\end{equation}

 By a \emph{localization scheme} is meat a procedure that 
relates the physical volume $P$ to its quantum degrees of freedom ${\cal H}_P$ by partitioning the total Hilbert space $\cH$ of the field into $\cH_P \otimes \cH_R$. As recalled in Sec. \ref{sec:2.1},
the partition $\cH = \cH_P \otimes \cH_R$ of a quantum system is induced by the choice of two commuting, and complete, subalgebras of operators acting on $\cH$~\cite{paolo2}. If $\x$ is a point in $P$ and $\y$ is not, i.e. $\y \in R$, then from the commutation relations \eqref{sec4:commu} we trivially have 
\begin{equation}
[\phi(\x),  \phi(\y)]\ =\
[\pi(\x),  \pi(\y)]\ =\
[\phi(\x),  \pi(\y)]\ =\ 0\, .
\end{equation}
Also linear combinations of $\phi$, $\pi$ and their spatial derivatives
commute if they belong to the two separate regions $P$ and ${R}$. 
In other words, relation \eqref{commute} is satisfied if we take as the algebra of operators
${\cal A}(P)$ the one generated by the local fields $\phi$ and $\pi$ with labels $\x$ in $P$. 
The corresponding partition is called the \emph{standard TPS} or the \emph{standard localization scheme}.

The non self-adjoint 
Newton-Wigner fields $a(\x)$ are just defined as the Fourier transform of $a_\kk$:
\begin{equation} \label{newton}
a(\x) = \frac{1}{(2 \pi)^{3/2}} \int d^3 k \, a_\kk\, e^{ i \kk \cdot \x} , \qquad
\ada(\x) = \frac{1}{(2 \pi)^{3/2}} \int d^3 k \, \ada_\kk\, e^{- i \kk \cdot \x} .
\end{equation}
Eq. (\ref{newton}) can be seen as a Bogoliubov transformation that doesn't mix creators with 
annihilators and therefore doesn't change the particle content of the system. As for any 
Bogoliubov transformation the commutation relations are preserved, i.e. 
$[a(\x), a(\xp)]=0,\,  [a(\x), \ada(\xp)]=\delta^3(\x - \xp)$.
As before, if $\p \in P$ and $\rr \in R$ 
(i.e. $\rr \notin P$), we have  
\begin{equation} \label{commu}
[a(\x),  a(\y)]\,=\
[a(\x),  \ada(\y)]\, =\ 0\, ,
\end{equation}
so that the subalgebras produced by the Newton Wigner fields also induce a TPS on $\cH$. 
The corresponding localization is called \emph{Newton-Wigner localization}.

The above defined operator $\ad(\x)$ is directly related to the Newton-Wigner (NW) position operator \cite{nw} in that, acting on the vacuum, it produces an eigenvector of eigenvalue $\x$. Note that the relativistic invariant measure $1/\sqrt{2 w_k}$ is absent from the integrand and therefore those operators are not relativistically invariant. 
In the definition of the local relativistic fields $\phi$, 
in which the invariant relativistic measure appears in the integral, namely:
\begin{equation} \label{field2}
\phi(\x) =  \frac{1}{(2 \pi)^{3/2}} \int \frac{d^3 k}{\sqrt{2 w_k}}
\left(a_\kk  e^{i \kk \cdot \x} + \ada_\kk e^{- i \kk \cdot \x}\right)\, .
\end{equation}

A particle localized in $\x$ (I have fixed the time at some fix $t$ throughout this section) is described in the two schemes by two different states of the theory: 
\begin{eqnarray} \label{standardlocalized}
|\x, {\rm standard}\ket & = & \phi(\x) |0\ket ,\\
|\x, {\rm NW}\ket & = & \ad(\x) |0\ket . 
\end{eqnarray}
Although the above is the general interpretation, the concept of localization of a particle is very severely challenged in the standard localization scheme (see Sec. \ref{sec:loca} below), while is not in the NW-scheme.

NW-localization is not a relativistic invariant property because a particle perfectly NW-localized according  to some observer is instead ``spread" when described by a boosted one \cite{stefa}. On the other hand, the dynamics is still relativistically invariant; changing localization does not modify the Hamiltonian nor the other generators of the Poincar\'e group.   We refer to the extensive literature for more technical details (e.g. \cite{nw,stefa,margaret}) and philosophical implications \cite{flem} of NW operators.
Finally, I would like to stress again that changing the tensor product structure $\cH_P \otimes \cH_R$, i.e. changing localization scheme, is not like moving the points
of space across the border of $P$, or choosing some different smearing or compact support
function for our definitions. As showed in Sec. \ref{sec:2.1} with the extreme example of a lattice made by only two-points, changing TPS is deeper than ``playing with the parts'' of a set.

\subsection{Some (very intuitive) properties of NW localization}
 
 In this subsection I am going to recall some properties of NW-localization. The main draw-back of this localization is its not being Lorentz invariant in a sense that I have already mentioned. The local relativistic fields $\phi(x, t)$ on the other hand, do not transform under Lorentz and are therefore usually considered as the good variable to be associated to the point/events $(x,t)$. However, the type of description that I am proposing here is based, from the very beginning, on some chosen  splitting \emph{space+time}. Regions of space can in fact be described as subsystems only after such a splitting has been chosen. Regardless of the localization scheme, Lorentz invariance is still a property of the \emph{dynamics}: the total Hamiltonian, the momentum and boost operators are the same and satisfy the appropriate commutation relations. Moreover, Newton and Wigner's attempt~\cite{nw} to introduce a well-defined and well-behaved position operator in relativistic field theories is worth of serious consideration for several intuitive -- even trivial -- good properties that are not shared by the usual localization scheme.  Some of the interrelated properties that make NW-localization an attractive rationale to characterize the degrees of freedom of a region of space are summarized in this subsection.

\subsubsection{The vacuum is a product state and each region has its own Fock space}

A fundamental property, from which other important follow, comes the basic request by Newton and Wigner that a particle localized in $\x$ and one localized in $\xp$ be described by two orthogonal states in the theory. This is at the basis of the possibility of having different regions of space characterized by independent Fock structures.
The Hilbert space of the field theory has in fact a Fock structure:
  \begin{equation}
	\cH = {\mathbb C} \oplus \cH_1 \oplus\ldots \oplus \cH_n\oplus\ldots \;,
	\label{fock}
\end{equation}
where $\cH_1$ is the single particle space and the $n$-particles space, $\cH_n$, is given, for a scalar theory, by the symmetric tensor product of $n$ copies of $\cH_1$. \emph{Ladder operators} are operators that take a vector of $\cH_j$ into one of $\cH_{j+1}$. In every localization scheme, the algebra of operators defined in the region of space $P$ contains a ladder operator. 
According to the NW scheme, these are, for instance, the NW operators $\ad(\x)$ of eq. \eqref{newton}, with $\x \in P$, and their superpositions. In the standard formalism, on the opposite,
one can consider the negative energy part of \eqref{field2}:
\begin{equation} 
\phi^-(\x) =  \frac{1}{(2 \pi)^{3/2}} \int \frac{d^3 k}{\sqrt{2 w_k}}
\ad_\kk e^{- i \kk \cdot \p}\, 
\end{equation} 
and superpositions. 
By applying the ladder operators of ${\cal A}(P)$ and ${\cal A}(R)$ to the vacuum state, we find two linear 
varieties $P_1$ and $R_1$ in $\cH_1$, representing the one-particle excitations inside and outside 
$P$ according to some localization scheme. Accordingly, the single particle space $\cH_1$ 
decomposes into a direct sum,
\begin{equation} \label{direct}
	\cH_1=P_1\oplus R_1	\, .
\end{equation}
The key point here is that $P_1$ and $R_1$ are not necessarily orthogonal. They are orthogonal in NW because of the 
commutation relations \eqref{commu} but not in the standard localization scheme, since
the two-point function $\bra 0 |\phi(x) \phi(x')|0\ket$ doesn't vanish outside 
the lightcone. When $P_1$ and $R_1$ are orthogonal, one can make the useful identification
\begin{equation} \label{identification} 
P_1 \longrightarrow P_1 \otimes |0\ket_R, \qquad R_1 \longrightarrow |0\ket_P \otimes R_1,
\end{equation}
which, rather intuitively, means that a particle well localized inside $R$ leaves $R$ 
``empty'' and vice versa.
This is not possible if $P_1$ and $R_1$ are not orthogonal because the RHSs of \eqref{identification} 
are orthogonal by construction. 

When I refer to the properties of NW-localization as ``trivial" I mean that they are somewhat intuitive and expected: the presence of a particle in the region $P$ leaves $R$ in the vacuum. This extends to any number of particles: the intuitive interpretation being
``if I have $n$ particles they can be all in $p$ and leave $R$ 
empty, or I can have $n-1$ particles in $P$ and one particle in $R$, or $n-2$  etc\dots''. 
In formulae, the entire Fock space $\cH$ decomposes into two Fock spaces $\cH_P$ and $\cH_R$:
 \begin{equation} \label{fockdecomp}
	\cH = \bigoplus_{n=0}^{\infty}\cH_n=\bigoplus_{n=0}^{\infty}\bigoplus_{k=0}^{n}P_k\otimes R_{n-k} 
	= \bigoplus_{n,\,m=0}^{\infty}P_n\otimes R_m \equiv \cH_P\otimes\cH_R\;.
\end{equation} 
This is not true in the standard localization scheme, where the corresponding $\cH_P$ and $\cH_R$ are not, 
independently, Fock spaces.

Note that, in the NW case, $P_1$ and $R_1$ are orthogonal subspaces of $\cH_1$ that correspond precisely to the regions of space of first quantization (where they are in fact subspaces rather than subsustems!). Thus, in the NW scheme we can fairly interpret each volume as a subsystem with an internal Fock structure analogous to the global one. 
On the opposite, in the standard scheme, the state of a particle localized in $P$ is not orthogonal to that of a particle localized in $R$; as a consequence, particles are not separately defined in $\cH_P$ and $\cH_R$ (see also \cite{rov} on this). This is strictly related to the vacuum being entangled in the standard scheme. 

A direct corollary is that the vacuum is a product state in the NW-localization. This is another related notable difference between the two schemes on which I comment more thoroughly in Sec. \ref{sec:separable} below. Since the entanglement of the vacuum is a very well studied and understood property of the standard localization scheme,  a crucial issue is to understand whether or not such a quantity accounts for practically measurable correlations, i.e. whether or not the entanglement of the vacuum has any operational meaning.
If the procedure described in \cite{reznik} to create EPR pairs from
vacuum entanglement  turned out to be experimentally practicable, this would strongly suggest that the 
standard localization scheme is the correct way to isolate the localized quantum degrees of freedom.

\subsubsection{The Thermal Entropy of a region of space is a sound thermodynamic quantity}

Thermal entropy in the NW localization has been the subject of a paper \cite{sergio} to which I refer for more details. In the usual localization, the calculation of the entropy of a region of space is plagued by infinities proportional to the area of the region that cannot be renormalized by standard methods \cite{en3}, i.e., by local counterterms. In \cite{sergio} we considered, in a $(d+1)$-dimensional Minkowski space, a free scalar field in a thermal state, $\rho_{\rm thermal} \propto e^{- \beta H}$. We then 
calculated the entropy 
\begin{equation}
S \, =  -{\rm Tr}_P (\rho_P \ln \rho_P) .
\end{equation}
relative to the region of space $P$. In the above, the trace is taken over the system $P$ and the (non-normalized) density matrix $\rho_P$ is given by $\rho_P \propto {\rm Tr}_R\, e^{- \beta H}$ and $H$ is the Hamiltonian given in \eqref{sec4:ham}, just generalized to the $d+1$ dimensional case.
In~\cite{sergio} we showed that if the NW-localization is used, and if traces are taken accordingly, the result is free of divergences and in accordance with thermodynamics/classical statistics: each region of space is a well defined thermodynamic systems if the corresponding quantum degrees of freedom are isolated and defined according to the NW-prescription. 

The calculation exploits the property of NW localization according to which, as remarked earlier in this subsection, each region of space has its own Fock space-structure. 
In fact, for each of the two factors $\cH = \cH_P\otimes \cH_R$, we have
\begin{equation}
\cH_P =  {\mathbb C} \oplus P_1 \oplus P_2 \oplus \ldots \oplus P_n\oplus\ldots \;\qquad
\cH_R =  {\mathbb C} \oplus R_1 \oplus R_2 \oplus \ldots \oplus R_n\oplus\ldots \ .
\end{equation}
Therefore,  in order to calculate traces, one can sum on a basis in a Fock space of given particle number and then sum over all such Fock subspaces. Schematically, if I want to trace on the external region $R$, I find
\begin{equation}
{\rm Tr}_R\,  \cdot = \bra 0_R| \cdot |0_R\ket + \int_R d^d r  \bra\rr| \cdot |\rr\ket +
\int_R d^d r  d^d r' \bra\rr \rr'| \cdot |\rr \rr'\ket +\dots \, ,
\end{equation}
where $|0\ket = |0\ket_P\otimes |0\ket_R $ and $|\rr_1 \dots \rr_n\ket = \frac{1}{\sqrt{n!}} \ad(\rr_1) \dots \ad(\rr_n)|0\ket$. The (non normalized) reduced density matrix $\rho_P \propto {\rm Tr}_R\, e^{- \beta H}$
is block diagonal in each Fock subspace of given particle number. In order to deal with a non-renormalized density matrix we used the trick~\cite{en3} 
\begin{equation} \label{entropy}
S \, \equiv\,  -{\rm Tr}_P (\rho_P \ln \rho_P)\, = \, \left. \left(-\frac{d}{d n} + 1\right)\ln {\rm Tr}_P \, \rho_P^n \right|_{n=1} \, .
\end{equation}
In order to evaluate the RHS of the above expression, it is useful to note that 
the trace of its $n^{\rm th}$ power of $\rho_P$
nicely rearranges  in an exponential, giving 
\begin{equation} \label{exp}
{\rm Tr}_P \, \rho_P^n \ = \ \exp\left(\sum_{j=1}^{\infty}\frac{1}{j} {\rm Tr}\, K^{j n}\right)\, ,
\end{equation}
where $K$ is the two-point function
\begin{equation}
K(\x_1, \x_2)\, \equiv \frac{_P\bra 0| a(\x_1)\, \rho_P\, \ad(\x_2)|0\ket_P}{_P\bra 0|\, \rho_P \, |0\ket_P},
\end{equation}
where $a(\x)$ and $\ad(\x)$ are the Newton Wigner operators \eqref{newton} and $\x_1$ and $\x_2$ are points 
inside $P$. The trace on the RHS of \eqref{exp} is made inside subsystem $P$ 
and limited to one-particle subspace:
\begin{equation}
{\rm Tr}\, K^m \ \equiv \ \int_{\x_1 \dots \x_m \in P} dx_1 dx_2 \dots dx_m\, K(\x_1,\x_2)K(\x_2,\x_3) \dots K(\x_m,\x_1).
 \end{equation}
 Each of the above expression has also a nice diagrammatic form for which I refer the reader to~\cite{sergio}. The results of are summarized as follows. The high-temperature/large-volume limit perfectly agrees with the results from classical statistical mechanics and gives 
\begin{eqnarray}\label{hightentropy}
S_d= (V T^d) \frac {(d-1)!}{2^{d-1}\pi^{\frac d2} \Gamma(d/2)} (d+1) \zeta(d+1)+ {\cal O}(V T^d)^0 \ ,
\end{eqnarray}
 where $T = 1/\beta$ is the temperature of the chosen Gibbs state. 
 At low temperature/small volume, as for the usual approach after the subtraction of the infinite contribution, thermal entropy goes to zero, although with a different behavior. The only known explicit calculation in standard localization is in 1+1 dimension~\cite{cala} and in higher dimensions by using AdS/CFT~\cite{ads}. From those studies it turns out that in standard localization, after the subtraction, thermal entropy is sub-extensive, $S_{\rm therm}\simeq (VT^d)^{(d+1)/d}$ at low temperature, whereas our regularized entropy approaches extensivity from above. For small $VT^d$, we obtain 
\begin{equation} 
 S\simeq -VT^d \ln VT^d + {\cal O}(V T^d). 
 \end{equation}

\subsubsection{NW allows localized states, standard localization does not} \label{sec:loca}

The usual localization scheme seriously challenges any idea of \emph{localized state}.
In Sec. \ref{sec:4.1} I recalled that, according to the usual interpretation, a particle instantaneously created at time $t$ at point $\x$ is described by the state of the theory (eq. \ref{standardlocalized})
\begin{equation} \label{state}
|\x, t \ {\rm standard}\ket  =  \phi(\x,t) |0\ket .
\end{equation}
The problem with such a description is that the above state is different from the vacuum at any point $\x ' \neq \x$. In other words, the particle is not localized at all. There is, in fact, a  precise and intuitive sense in which a state is localized. One very natural definition is the following. A
 state $|\psi\ket$ is \emph{strictly localized} \cite{strict} outside of $P$ if, for any possible observable $A$ in ${\cal A}(P)$, $\bra\psi|A|\psi\ket=\bra 0|A|0\ket$. In other words, if we excite some 
degrees of freedom that are ``strictly localized'' outside $P$, 
the state of affairs inside $P$ is the same as the one 
of the vacuum.  It turns out
that no state with finite energy has this property in the standard localization scheme. The state \eqref{state} which is commonly described as ``a particle at position $\x$''
is in fact different from the vacuum in any region $P$ with $\x \notin P$, i.e. 
$\rho_P \equiv {\rm Tr}_R  |\x, t \ {\rm standard}\ket \bra \x, t \ {\rm standard} | \neq {\rm Tr}_R  |0\ket \bra 0|$.
This property, that can be traced back to Reeh-Schlieder theorem \cite{clift,terno}, is related, once again, 
with the fact that the vacuum is entangled in the standard scheme. On the other hand, low energy 
excitations can be ``strictly local'' in the NW scheme because of the factorization 
\eqref{identification} that leaves $P$ empty and in its ``local vacuum'' whenever we excite some degrees
of freedom somewhere else (i.e. in $R$).

\subsection{Again on vacuum entanglement vs separability} \label{sec:separable}

Entanglement is a  well understood -- and measurable -- property of many condensed matter systems in their ground state (see, e.g. \cite{vidal}). When standard localization is used, the vacuum in quantum quantum field theory shares the same property: different regions of space are entangled, with an entropy proportional to the area of the boundary between them~\cite{en1,en2,en3,en4,en5}.  One of the most popular interpretations of black hole entropy~\cite{damour} is in fact to consider it as the entanglement entropy of the quantized fields on the black hole background (e.g. ~\cite{jac,dvali}). The idea is to trace over the degrees of freedom inside the horizon because they are not accessible from the outside. This gives a UV-divergent answer proportional to the area of the boundary, which is compatible with black hole entropy if the cut-off is suitably chosen. The question that I am going to raise here is whether those correlations that vacuum entropy is accounting for have any operational meaning, i.e., whether they can be used as a source of entanglement in EPR-like experiments. In the black hole case, because of the presence of the event horizon, the correlations between inside and outside are argued to be lost, which should explain the entropy of the black hole itself. But are vacuum correlations ever really accessible anyway, for instance, in flat space? 
In this section I am going to argue for a negative answer to that question: the ground state correlations are accessible in a condensed matter system, but the vacuum state correlations are not in QFT.

\subsubsection{Samantha and Vincent}

Consider a quantum spin system with a general nearest-neighbour Hamiltonian. Schematically,
\begin{equation} \label{ham}
H \, = \, \sum_{\bf j} f(\sigma_{\bf j}) + \sum_{\bf j} \sum_{{\bf k} \in {\bf j}} g(\sigma_{\bf j} \otimes \sigma_{{\bf k}})\,  +\,  \dots \ ,
\end{equation}
where $\sigma_j$ are the local degrees of freedom, $f$ and $g$ are polynomials, internal indexes and more complicate tensorial structures have been omitted. The symbol $\in$ here stands for ``is neighbor of" and the dots possibly stand for next-to-nearest neighbor- or more than two spins- interactions.    The above Hamiltonian aims to resemble  a cut-off field theory, $\sigma_{\bf j}$ being the relativistic fields and the nearest neighbors interaction a lattice version of the gradient terms. 

Say that there exists a concrete realization of that spin system, and that such an apparatus is physically placed on the table of Samantha, an experimental quantum computer scientist. Samantha is very skilled and has enough instruments in her lab to have full access to the individual spins $\sigma_j$, she can pick one up and take it out of the system, she can make individual measurements etc\dots 
Now let us suppose that the dynamics internal to such a condensed matter system is extremely rich, say that (\ref{ham}) is a lattice version of the Standard Model or some lower-energy truncation of it. Then, if the system is big enough, there might be something similar to our world within it: some eigenstates of (\ref{ham}) will correspond to stable elements, more complicate configurations will correspond to scattering experiments. 
There will be quantum states, evolving according to (\ref{ham}), that we should be authorized to interpret as ``measurements" going on within the system. Finally, we will have observers! Call Vincent one of those observers. Vincent is made by the excitations of (\ref{ham}) as much as we are made of atoms, molecules etc \dots  

It is usually the case that the Hamiltonian (\ref{ham}) has an entangled ground state. More precisely, the ground state is entangled with respect to the quantum partition defined by the spins $\sigma_j$, which corresponds to the standard localization. For Samantha such an entanglement has a very clear operational meaning. As we have assumed, she has full access to the spins' degrees of freedom. Therefore, if she can manage to cool the system down and bring it close to its ground state, then she can divide the spins into two groups and measure the Bell correlations between them. Alternatively, she can take few spins out of the chain and see whether the state of the system rearranges afterwards. 

For Vincent the situation is different since (\ref{ham}) is his most fundamental -- field theory -- description. Typical objects in Vincent's world are not localized in Samantha's sense because they are eigenstates or quasi-eigentstates configurations of (\ref{ham}). This is discussed in Sec. \ref{sam} below and in~\cite{fabio2} in more detail. In quantum mechanics we are used to prepare states in arbitrary configurations. 
However, what seems peculiar of field theory at low density and big volume is that generic states generally evolve by radiating away decay products  until we are left, in a sufficiently large region of space, with field configurations which are approximate eigenstates of the total Hamiltonian: stable particles, molecules, crystals etc\dots 
This applies even more dramatically to strongly interacting theories: scattering products are known to hadronize very rapidly if they are not QCD eigenstates. 

We are arguing that Vincent does not have access to the spin degrees of freedom and, therefore, to the ground state correlations. The objects in his world (particles, atoms  etc\dots) are approximate eigenstates-excitations of \eqref{ham}. Whenever he measures quantum correlations he is doing so among such objects, not among the original spins. For Vincent, the most natural description in terms of subsystems effectively implies a different TPS than the one given by the degrees of freedom $\sigma$. He will be clever enough to reconstruct the Hamiltonian \eqref{ham} that successfully reproduces his experiments and note that such an Hamiltonian is \emph{local} (first neighbor interactions) with respect to some spin basis. But whenever he tries to operationally define detailed spacetime relations between his objects, instruments etc\dots he is effectively using a different basis. Effectively, Samantha and Vincent have two different fine-grained descriptions of spacetime. 

\subsubsection{Extracting the correlations from the vacuum; particle detectors} \label{sam}

What I just argued can be proven wrong straight away by an experiment of vacuum entanglement distillation. 
Such an experiment has in fact been conjectured, featuring two Unruh-DeWitt detectors. Those are objects with local interactions, i.e., Hamiltonians of the type
\begin{equation} \label{unruh}
H_{\rm Unruh\ DeWitt} \ =\ \lambda {\hat o}\, \phi(\x(t),t),
\end{equation}
where $\lambda$ is a dimensionless coupling and $\hat o$ is some operator internal to the detector. We can take, for simplicity, a two level detector, for which, ${\hat o} = {\hat o}_{\uparrow} + {\hat o}_{\downarrow}$, where ${\hat o}_{\uparrow} = |E\ket \bra0|$ and ${\hat o}_{\downarrow} = ({\hat o}_{\uparrow})^\dagger$ and $|0\ket$ and $|E\ket$ are the ground and excited states of the detector.

The ``distillation" of vacuum entanglement proposed in~\cite{reznik} relies on the \emph{dark counts} that those detectors are known to undergo in the vacuum~\cite{rabi}. Two detectors are kept at rest in Minkowski vacuum at a given relative distance and are both switched on and off at space-like separated events and for time intervals $\Delta t \ll 1/\Delta E$, $\Delta E$ being the energy gap inside the detectors. Through the interaction with the vacuum the detectors get entangled between each other and the above described dark counts should show EPR-like correlations. If ever realized, this experiment would give a precise operational meaning to the entanglement of the vacuum between different regions of space. 

The experiment proposed in~\cite{reznik} highlights a consistency within the standard localization scheme: the Unruh-DeWitt detector is localized in the standard sense and, consistently, ``feels" the vacuum correlations that link regions of space according to standard localization. 
In~\cite{fabio2} we have argued that realistic detectors have a more trivial response on the vacuum and do not perform \emph{dark counts} of such a fundamental origin. The reason, again, is that a detector in its ground state, and in the absence of quanta to detect, is represented in the full underlying field theory by an eigenstate of the total Hamiltonian. In fact, being an eigenstate of the Hamiltonian seems incompatible~\cite{fabio2,terno} with having local QFT interactions on the basis of the Reeh-Schlieder theorem~\cite{reeh}.

 In~\cite{fabio2} we have studied a simple local relativistic field theory with three scalar fields, $\chi$, $\eta$ and $\phi$ with masses $M$, $\mb$ and $m$ respectively and interaction Lagrangian ${\cal L}^I(x) \sim -\mu\, \chi(x)\eta(x)\phi(x)$. The coupling $\mu$ is massive and is assumed to be smaller than the masses of the other fields in order to allow a perturbative treatment. Moreover, $m\ll M<\mb$. The theory contains two stable particles (corresponding to the fields $\chi$ and $\phi$, of masses $M$ and $m$ respectively), and a meta-stable one ($\eta$, of mass $\mb$). We take the particle $\chi$ to describe our detector in the ground state. Such a particle can capture a light quanta $\phi$ and form the meta-stable state $\eta$, corresponding to our excited detector. To the stable particle $\chi$, as expected, correspond energy eigenstates of the full theory. This is shown explicitly in ~\cite{fabio2} by using a dressed particle formalism at first order in perturbation theory.  Then, we derived an effective model of detector (at rest) where the stable particle and the quasi-stable configurations correspond to the two internal levels, ``ground" and ``excited", of the detector, and the transitions rate of the original field theory are faithfully reproduced. The detector model (``the Frog",~\cite{fabio2}) reads:
\begin{equation} 
\label{ours}
		H_{\rm Frog} = \frac{\mu}{M}\left({\hat o}_{\uparrow} \Phi^+(\x) + {\hat o}_{\downarrow} \Phi^-(\x)\right),
\end{equation}
where, again, ${\hat o}_{\uparrow}$, ${\hat o}_{\downarrow}$ are the raising and lowering operators of the two level detector and the energy gap inside the detector is $\Delta E = \Delta M = \mb - M$. The complex fields $\Phi^+(\x) $  and $\Phi^-(\x)$ are defined in terms of the some annihilators produced by the ``dressing" and which are a combination of the degrees of freedom of the three fields we started with.

Although derived from a perfectly local field theory, our effective detector does not couple locally to the field $\phi$ but only, separately, to the positive and negative frequencies, $\Phi^-(\x)$ and $\Phi^+(\x)$, of a scalar field $\Phi(x) \equiv \Phi^-(\x) + \Phi^+(\x)$, which is a ``dressed" mixture of the original field $\phi$ and the other fields. The effective coupling~\eqref{ours} is non-local, because reproduces the behavior of a real-dressed, rather than bare, particle in the theory. Our detector \eqref{ours} does not click in the vacuum, it does click only in the presence of a quantum.
As a corollary, our analysis suggests that the experiment envisioned in~\cite{reznik} is not doable with real particle detectors, and that the entanglement of the vacuum has no operational implications and cannot possibly be used to create testable correlations\footnote{Correlations seem to be lost also in the presence of a coarse graining~\cite{massi}.}.

\subsection{Discussion}

One may argue that standard localization is just intrinsic to the formalism and that going for alternatives would imply to contradict quantum field theory itself. In fact, in some approaches to QFT, the degrees of freedom $\phi$ and $\pi$ are defined \emph{ab initio} as the localized objects of the theory. It is interesting, in this respect, to make a comparison with the more constructive approach e.g. of the beautiful book \cite{weinbook}. There, the entire QFT formalism is built step by step in order to give account for scattering experiments and decay processes; rather then a fine-grained spacetime description, the aim is to find amplitudes relating the asymptotic states, which are what is observable in scattering experiments. The relativistic fields $\phi$, as objects of the theory, are introduced at a later stage (Chap. 5) with the purpose of writing -- automatically and easily -- Lorentz invariant interactions. Those are in fact guaranteed if the Hamiltonian is local, in the usual sense, in the fields $\phi$, $\pi$.  

When I discuss alternative localization schemes I am not calling into question the local form of the Hamiltonian: in the case under consideration considered in this section, for example, the operator (\ref{ham}) does have the local form
\begin{equation} \label{local}
H =\frac{1}{2} \int d^3 x \left(\pi^2 + \nabla \phi^2 + m^2 \phi^2\right)
\end{equation}
when written in terms of $\phi$ and $\pi$.  The same operator can also be written in terms of the NW-operators by writing the Hamiltonian in terms of creators and annihilators as in \eqref{sec4:ham} and inverting the two Fourier transforms \eqref{newton},
\begin{equation} \label{hamnewton}
H \ = 
\int  d^3x \, d^3y \, K(|{\vec x} - {\vec y}|) \, a^\dagger({\vec x}) a({\vec y}), 
\end{equation}
where the function $K$ is a kernel that 
dies off as $K \sim e^{- m|{\vec x} - {\vec y}|}$ 
for $|{\vec x} - {\vec y}|\gg m^{-1}$ where $m$ is the mass of the field and $K \propto |{\vec x} - {\vec y}|^{-4}$ in the massless case.

Independently of the localization scheme, \eqref{local} and \eqref{hamnewton} is just the same operator that uniquely sets the dynamics among asymptotic states (cross sections, decay rates etc\dots). What \emph{is} dependent on the localization scheme is the fine-grained spacetime description. Take, for instance, the question: ``what's the average energy inside this room now?". According to the usual prescription, that corresponds to a QFT operator of the type
\begin{equation}
H_{\rm room}^{\rm standard}\  \simeq \ V \, \langle T^{00}\rangle_{\rm volume\ average} \ \simeq \ \frac{1}{2} \int_{\x \in \rm room} \!\!\!\!\!d^3 x \left(\pi^2 + \nabla \phi^2 + m^2 \phi^2\right)\  ,
\end{equation}
where the average symbol is a spatial average and $V$ is the volume of the submanifold ``room". 
In the NW localization the answer to the same question is given by a \emph{different} operator. When limited to the subsystem ``room", (\ref{hamnewton})
becomes instead
\begin{equation} \label{nwhamiltonian}
H_{\rm room}^{\rm NW} \ \simeq \ V \, \langle T^{00}\rangle_{\rm volume\ average} \ \simeq \
\int_{\x, \y \in \rm room} \!\!\!\!\! d^3x \, d^3y \, K(|{\vec x} - {\vec y}|) \, a^\dagger(\x) a(\y) .
\end{equation}
Alternatively, one can extend one of the two variables $\x$, $\y$ to the whole space, thereby considering also the interaction (inside/outside) part of the energy.
Note that the operators $H_{\rm room}^{\rm standard}$ and $H_{\rm room}^{\rm NW}$ coincide in the limit when the size of the room is much bigger that the Compton wavelength $m^{-1}$ of the field. For a  massless field the two localizations converge when the room is much bigger than the inverse temperature $T^{-1}$  of the quantum state considered. 

In summary,
as long as we are concerned only with the internal dynamics of the fields,
all TPSs describe precisely the same state of affairs: we are just considering different -- equally valid --
partitions into subsystems of the field system,
not changing its dynamics (cross sections, decay rates etc\dots). Things may possibly be different when also gravity is 
taken into account. Eq. \eqref{nwhamiltonian} suggests a brutally non-local coupling gravity--matter fields and, effectively, a smearing at small scales of the energy content of a region of space. If we stick with the genuine idea that (semiclassical) gravity is really just the geometry of the physical spacetime, then it is  crucial to understand how the matter degrees of 
freedom feeding into Einstein equations are ``localized'' in the physical spacetime
itself. By incorporating alternative localization schemes, semiclassical gravity inherits from the matter fields a certain amount of non-locality (see eq. \ref{nwhamiltonian}).

\section{Conclusions}

At the beginning of this review I listed few problems affecting the established low-energy theoretical framework for gravity. As opposed to other famous examples in the history of physics, the problems at hand do not seem to be manifest striking inconsistencies and, therefore, they are not guaranteed to be triggering any imminent major revision or breakthrough. However, in parallel with more standard approaches (model building with scalar fields, new -- local or non-local -- terms in the Lagrangian, string-inspired scenarios, extra dimensions etc\dots) it is also worth examining the very premises of such an established framework, questioning the parts that are believed to be true but still lack of experimental verification, trying substantially new directions.  More radical approaches, when not manifestly inconsistent, are still inevitably tentative, incomplete, hinting to future work.  

The directions mentioned in this review have semiclassical gravity as a starting point and are both based on the idea (Sec. \ref{sec:2}) that a deeper and more general description of spacetime is that of considering it as a quantum system -- containing the matter fields' degrees of freedom. Accordingly, regions of space at a given time are, very generally, quantum subsystems, rather than pieces of a manifold. In section~\ref{sec3} I explored the possibility that the metric manifold description is in fact just a small distance approximation; while doing this, I attempt to address directly some of the problems of the standard low-energy framework. As a guide for such a modification, I propose to consider an ultra-strong version of the equivalence principle, that I formulate and motivate in Sec. \ref{sec:3.3}. In section~\ref{sec4} I reviewed the advantages of considering a different localization scheme, i.e., a different rationale to associate to a region of space its appropriate degrees of freedom. This suggests a fine-grained spacetime description that differs from the standard one at scales comparable with the Compton wavelength of the fields and/or the temperature of the state considered.

\section*{Acknowledgments}

Niayesh Afshordi, Michele Arzano, Sergio Cacciatori, Denis Comelli, Fabio Costa, Olaf Dreyer, Justin Khoury,  Alioscia Hamma, Christian Marinoni, Savvas Nesseris, Maxim Pospelov, Constantinos Skordis, Lee Smolin, Eugene Stefanovich, Andrew Tolley, Shinji Tsujikawa, Filippo Vernizzi, Alberto Zaffaroni and Paolo Zanardi is a by far incomplete list of colleagues to whom I am indebted  for insights and/or collaborations on these topics.  Research at the Perimeter Institute is supported in part by the Government of Canada through NSERC and by the Province of Ontario through the Ministry of Research \& Innovation.\\

\end{document}